\documentclass[
 reprint,twocolumn,
 amsmath,amssymb,
 aps,
]{revtex4}

\usepackage{epsfig,bm,epsf,graphics}
\usepackage{graphicx}% Include figure files
\usepackage{dcolumn}% Align table columns on decimal point
\usepackage{bm}% bold math
\begin{document}
%\draft
\preprint{APS/123-QED}

\title{Quantum Theory of Helimagnetic Thin Films}% Force line breaks with \\

\author{H. T. Diep}
\email{diep@u-cergy.fr}
% \altaffiliation[Also at ]{Asia Pacific Center for Theoretical Physics,
% Hogil Kim Memorial Bldg. \#501, POSTECH San 31 Hyoja-dong, Nam-gu, Pohang Gyeongbuk 790-784, Korea.}%Lines break automatically or can be forced with \\
 \affiliation{%
Laboratoire de Physique Th\'eorique et Mod\'elisation,
Universit\'e de Cergy-Pontoise, CNRS, UMR 8089\\
2, Avenue Adolphe Chauvin, 95302 Cergy-Pontoise Cedex, France.\\
 }%


%\affiliation{
%\affiliation{
%\affiliation{%



\begin{abstract}

We  study properties of a helimagnetic thin film with quantum Heisenberg spin model by using the Green's function method.
Surface spin configuration is calculated by minimizing the spin interaction energy.  It is shown that the angles
between spins near the surface are strongly modified with respect to the bulk configuration.
Taking into account this surface spin reconstruction, we calculate self-consistently the
spin-wave spectrum and the layer magnetizations as  functions of temperature up to the disordered phase.
The spin-wave spectrum shows the existence of a surface-localized branch which causes a low surface magnetization.
We show that quantum fluctuations give rise to a crossover between the surface magnetization and interior-layer magnetizations
at low temperatures.  We calculate  the transition temperature and show that it depends strongly on the helical angle.
Results are in agreement with existing experimental observations on the stability of helical structure in thin films and on the insensitivity
of the transition temperature with the film thickness.
We also study effects of various parameters such as surface exchange and anisotropy interactions. Monte Carlo simulations for the classical spin model are also carried out for comparison with the quantum theoretical result.
\begin{description}
\item[PACS numbers: 75.25.-j ; 75.30.Ds ; 75.70.-i ]
\end{description}
\end{abstract}

\pacs{Valid PACS appear here}% PACS, the Physics and Astronomy
\maketitle

%\tableofcontents

\section{Introduction}
Helimagnets have been discovered a long time ago by Yoshimori \cite{Yoshimori} and Villain \cite{Villain59}.  In  the simplest model,
the helimagnetic ordering
is non collinear due to a competition between nearest-neighbor (NN)  and next-nearest-neighbor (NNN) interactions:
for example, a spin in a chain turns an angle $\theta$ with respect to its previous neighbor.
Low-temperature properties in helimagnets such as spin-waves \cite{Harada,Rastelli,Diep89,Quartu1998} and
heat capacity \cite{Stishov} have been extensively investigated. Helimagnets belong to a class of frustrated vector-spin systems.
In spite of their long history, the nature of the phase transition in bulk helimagnets as well as in other non collinear magnets such as stacked
triangular XY and Heisenberg antiferromagnets has been elucidated only recently \cite{Diep89b,Ngo08,Ngo09}. For reviews on
many aspects of frustrated spin systems, the reader is referred to Ref. \onlinecite{DiepFSS}.

In this paper, we study a helimagnetic thin film with the quantum Heisenberg spin model. Surface effects in thin films have
been widely studied theoretically, experimentally and numerically, during the last three decades \cite{Heinrich,Zangwill}. Nevertheless, surface effects in helimagnets have
only been recently studied: surface spin structures \cite{Mello2003}, Monte Carlo (MC) simulations \cite{Cinti2008} and a few experiments \cite{Karhu2011,Karhu2012}. Helical magnets present potential applications in spintronics with predictions of spin-dependent electron transport in these magnetic materials \cite{Heurich,Wessely,Jonietz}.  This has motivated the present work.
We shall use the Green's function (GF) method to study a quantum spin model on a helimagnetic thin film of body-centered cubic (BCC) lattice.  The GF method has been initiated
by Zubarev \cite{zu} for collinear bulk magnets (ferromagnets and antiferromagnets) and by Diep-The-Hung {\it et al.} for collinear
surface spin configurations \cite{Diep1979}.  For non collinear magnets, the GF method has also been developed for bulk
helimagnets \cite{Quartu1998} and for frustrated films \cite{NgoSurface,NgoSurface2}.  In helimagnets, the presence of a surface modifies the competing forces acting on surface spins. As a consequence, as will be shown below, the angles between neighboring spins become non-uniform, making calculations harder. This explains why there is no microscopic calculation so far for helimagnetic films.

The paper is organized as follows. In section II, the model is presented and classical ground state (GS) of the helimagnetic film is determined. In section III, the general GF method for non-uniform spin configurations is shown in details. The GF results are shown in section IV where the spin-wave spectrum, the layer magnetizations and the transition temperature are shown. Effects of surface interaction parameters and the film thickness are discussed. Concluding remarks are given in section V.

\section{Model and classical ground state}\label{GSSC}

Let us recall that bulk helical structures are due to the competition of various kinds of interaction \cite{Yoshimori,Villain59,Bak,Plumer,Maleyev}. We consider hereafter the simplest model for a film: the helical ordering is along one direction, namely the $c$-axis perpendicular to the film surface.

We consider a thin film of BCC lattice of $N_z$ layers, with two symmetrical surfaces perpendicular to
the $c$-axis, for simplicity.  The exchange Hamiltonian is given by
\begin{equation}
\mathcal H_e=-\sum_{\left<i,j\right>}J_{i,j}\mathbf S_i\cdot\mathbf
S_j  \label{eqn:hamil1}
\end{equation}
where $J_{i,j}$ is the interaction between two quantum Heisenberg spins $\mathbf S_i$ and $\mathbf S_j$ occupying the lattice sites $i$ and $j$.

To generate a bulk helimagnetic structure, the simplest way is to take a ferromagnetic interaction between NNs, say $J_1$ ($>0$),
and an antiferromagnetic interaction between NNNs,  $J_2<0$.  It is obvious that if $|J_2|$ is smaller than a critical value $|J_2^c|$,
the classical GS spin configuration is ferromagnetic \cite{Harada,Rastelli,Diep89}.
Since our purpose is to investigate the helimagnetic structure near the surface and surface effects, let us consider the case of a
helimagnetic structure only in the $c$-direction perpendicular to the film surface. In such a case, we assume a non-zero $J_2$ only on the $c$-axis.
This assumption simplifies formulas but does not change the physics of the problem since including the uniform helical angles in two other
directions parallel to the surface will not introduce additional surface effects.  Note that the bulk case of the above
quantum spin model have been studied by the Green function method \cite{Quartu1998}.

\begin{figure}[htb]
\centering
\includegraphics[width=5cm]{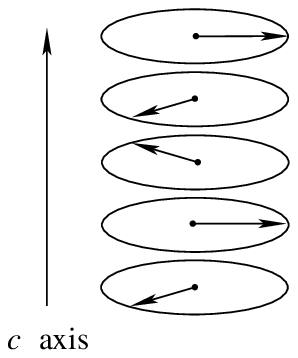}  % .eps
\includegraphics[width=7cm,angle=0]{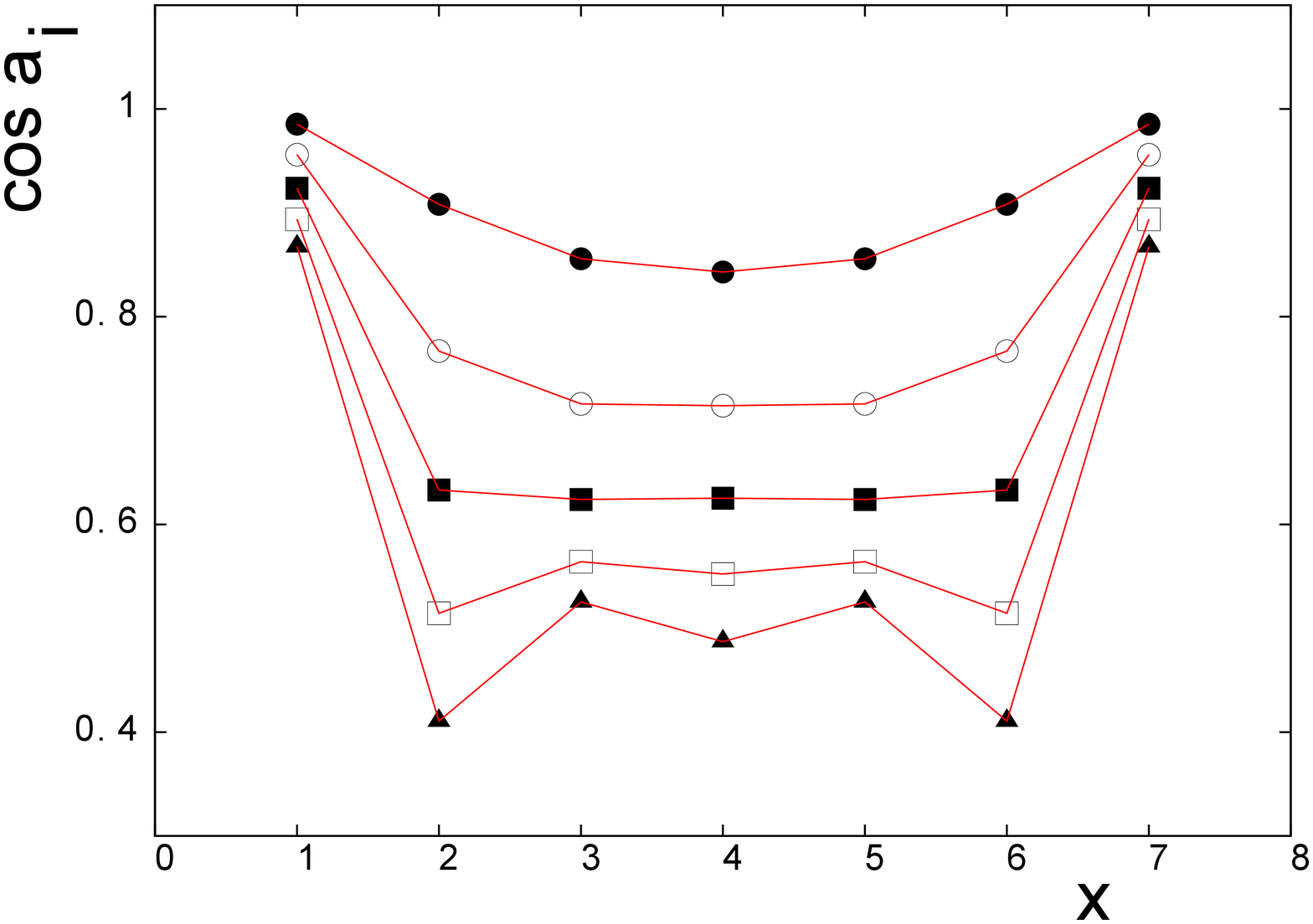}  % .eps
\caption{ Top: Bulk helical structure along the $c$-axis,
in the case $\alpha=2\pi/3$, namely $J_2/J_1=-2$.  Bottom: (color online) Cosinus of $\alpha_1=\theta_1-\theta_2$, ..., $\alpha_7=\theta_7-\theta_8$ across the film
for $J_2/J_1=-1.2,-1.4,-1.6,-1.8, -2$ (from top) with $N_z=8$: $a_i$ stands for $\theta_i-\theta_{i+1}$ and $x$ indicates the film layer $i$ where the angle $a_i$ with the layer $(i+1)$ is shown.  The values of the angles are given in Table I: a strong rearrangement of spins near the surface is observed. }\label{GSA}
\end{figure}

Let us recall that the helical structure in the bulk is planar: spins lie in planes perpendicular to the $c$-axis: the angle between two NNs in
the adjacent planes is a constant and is given by $\cos \alpha=-J_1/J_2$ for a BCC lattice. The helical structure exists therefore if $|J_2|\geq J_1$,
namely $|J_2^c|$(bulk)$=J_1$ [see Fig. \ref{GSA} (top)]. To calculate the classical GS surface spin configuration, we write down the expression of
the energy of spins along the $c$-axis, starting from the surface:
\begin{eqnarray}
E&=& -Z_1 J_1 \cos (\theta_1-\theta_2)-Z_1 J_1 [\cos (\theta_2-\theta_1)\nonumber\\
&&+ \cos (\theta_2-\theta_3)]+...\nonumber\\
&&-J_2 \cos (\theta_1-\theta_3)-J_2 \cos (\theta_2-\theta_4)\nonumber\\
&&-J_2[\cos (\theta_3-\theta_1)+ \cos (\theta_3-\theta_5)]+...\label{EC}
\end{eqnarray}
where $Z_1=4$ is the number of NNs in a neighboring layer, $\theta_i$ denotes the angle of a spin in the $i$-th layer
made with the Cartesian $x$ axis of the layer. The interaction energy between two NN spins in the two adjacent layers $i$ and $j$
depends only on the difference $\alpha_{i}\equiv \theta_i-\theta_{i+ 1}$. The GS configuration corresponds to the minimum of $E$.
We have to solve the set of equations:
\begin{equation}
\frac{\partial E}{\partial \alpha_i}=0, \ \ \ \mbox{for}\ \ i=1,N_z-1
\end{equation}
Explicitly, we have
\begin{eqnarray}
\frac{\partial E}{\partial \alpha_1}&=&8J_1\sin \alpha_1+2J_2\sin (\alpha_1+\alpha_2)=0\label{A1}\\
\frac{\partial E}{\partial \alpha_2}&=&8J_1\sin \alpha_2+2J_2\sin(\alpha_1+\alpha_2)\nonumber\\
&&+2J_2\sin(\alpha_2+\alpha_3)=0\label{A2}\\
\frac{\partial E}{\partial \alpha_3}&=&8J_1\sin \alpha_3+2J_2\sin(\alpha_2+\alpha_3)\nonumber\\
&&+2J_2\sin(\alpha_3+\alpha_4)=0\label{A3}\\
\frac{\partial E}{\partial \alpha_4}&=&... \nonumber
\end{eqnarray}
where we have expressed the angle between two NNNs as follows: $\theta_1-\theta_{3}=\theta_1-\theta_{2}+ \theta_2-\theta_{3}=\alpha_1+\alpha_2$ etc.
In the bulk case, putting all angles $\alpha_i$ in Eq. \ref{A2} equal to $\alpha$ we get $\cos \alpha=-J_1/J_2$ as expected.
For the spin configuration near the surface,  let us consider in the first step only three parameters $\alpha_1$
(between the surface and the second layer), $\alpha_2$ and $\alpha_3$. We take $\alpha_n=\alpha$ from $n=4$ inward up
to $n=N_z/2$, the other half being symmetric.
Solving the first two equations, we obtain
\begin{equation}\label{angle}
\tan \alpha_2=-\frac {2J_2(\sin \alpha_3+\sin \alpha_1)}{8J_1+2J_2(\cos \alpha_3+\cos \alpha_1)}
\end{equation}
The iterative numerical procedure is as follows: i)
replacing $\alpha_3$ by $\alpha=\arccos  (-J_1/J_2)$ and solving (\ref{A1}) and (\ref{angle}) to obtain
$\alpha_1$ and $\alpha_2$, ii) replacing these values into
(\ref{A3}) to calculate $\alpha_3$, iii) using this value of $\alpha_3$ to solve again (\ref{A1}) and (\ref{angle}) to obtain new values of $\alpha_1$ and $\alpha_2$,
iv) repeating step ii) and iii) until the convergence is reached within a desired precision.    In the second step, we use
 $\alpha_1$,  $\alpha_2$ and  $\alpha_3$ to calculate by iteration  $\alpha_4$, assuming  a bulk value for $\alpha_5$. In the third step, we use  $\alpha_i$ ($i=1-4)$ to calculate  $\alpha_5$ and so on.
The results calculated for various $J_2/J_1$ are shown in Fig. \ref{GSA} (bottom) for a film of $N_z=8$ layers.
The values
obtained are shown in Table \ref{table}.  Results of $N_z=16$ will be shown later.

\begin{widetext}
\begin{center}
\begin{table}
\begin{tabular}{|l|c|c|c|c|r|}
\hline
$J_2/J_1$ & $\cos \theta_{1,2}$ & $\cos\theta_{2,3}$ & $\cos\theta_{3,4}$ & $\cos\theta_{4,5}$ & $\alpha$(bulk)  \\
\hline
&&&&&\\
-1.2  &  0.985($9.79^\circ$) &  0.908($24.73^\circ$)       &    0.855($31.15^\circ$)    &   0.843($32.54^\circ$)  &   $33.56^\circ$    \\
-1.4 &  0.955($17.07^\circ$)&  0.767($39.92^\circ$)  &  0.716($44.28^\circ$)    &   0.714($44.41^\circ$)   &  $44.42^\circ$     \\
-1.6 & 0.924($22.52^\circ$) &  0.633($50.73^\circ$) & 0.624($51.38^\circ$)  &  0.625($51.30^\circ$)   &  $51.32^\circ$        \\
-1.8     &  0.894($26.66^\circ$)  &  0.514($59.04^\circ$)  & 0.564($55.66^\circ$)   &  0.552($56.48^\circ$)  &  $56.25^\circ$    \\
-2.0       &  0.867($29.84^\circ$)  &  0.411($65.76^\circ$) &  0.525($58.31^\circ$)   &  0.487($60.85^\circ$)  & $60^\circ$   \\
&&&&&\\
\hline
\end{tabular}
\caption{Values of $\cos \theta_{n,n+1}=\alpha_n$ between two adjacent layers are shown for various
values of $J_2/J_1$. Only angles of the first half of the 8-layer film are shown: other angles are, by symmetry,
$\cos\theta_{7,8}$=$\cos\theta_{1,2}$, $\cos\theta_{6,7}$=$\cos\theta_{2,3}$, $\cos\theta_{5,6}$=$\cos\theta_{3,4}$. The values in parentheses are angles in degrees.
The last column shows the value of the angle in the bulk case (infinite thickness).
For presentation, angles are shown with two  digits. \label{table} }
\end{table}
\end{center}
\end{widetext}

\vspace{1cm}

Some remarks are in order: i) result shown is obtained by iteration with errors less than $10^{-4}$ degrees,  ii) strong angle variations  are observed near the surface with oscillation for strong $J_2$, iii) the angles at the film center are close to the bulk value $\alpha$ (last column), meaning that the surface reconstruction affects just a few atomic layers, iv) the bulk helical order is stable just a few atomic layers away from the surface even for films thicker that $N_z=8$ (see below). This helical stability has been experimentally observed in holmium films \cite{Leiner}.

Note that using the numerical steepest descent method described in Ref. \onlinecite{NgoSurface} we find the same result.

In the following, using the spin configuration obtained at each $J_2/J_1$ we calculate the spin-wave excitation and properties
of the film such as the zero-point spin contraction, the layer magnetizations and the critical temperature.

\section{Green's function method}

\begin{figure}[htb]
\centering
\includegraphics[width=7cm]{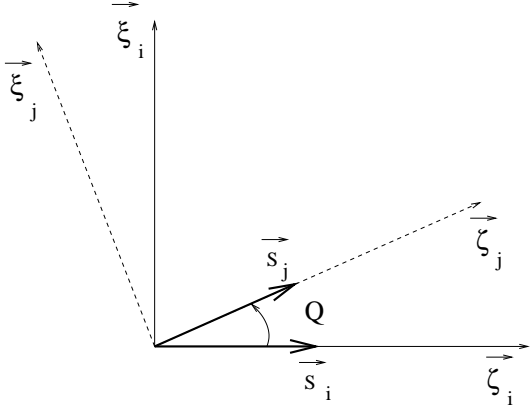}  % .eps
\caption{ Local coordinates in a $xy$-plane perpendicular to the $c$-axis. $Q$ denotes $\theta_{j}-\theta_i$.}\label{local}
\end{figure}

Let us define the local spin coordinates as follows:  the quantization axis of spin $\vec S_i$ is on
its $\zeta_i$ axis which lies in the plane, the $\eta_i$ axis of $\vec S_i$ is along the $c$-axis, and the $\xi_i$
axis forms with $\eta_i$ and $\zeta_i$ axes a direct trihedron. Since the spin configuration is planar, all spins
have the same $\eta$ axis. Furthermore,  all spins in a given layer are parallel.  Let $\hat \xi_i$, $\hat \eta_i$
and $\hat \zeta_i$ be the unit vectors on the local $(\xi_i,\eta_i,\zeta_i)$ axes. We write
\begin{eqnarray}
\vec S_i&=&S_i^x\hat \xi_i+S_i^y\hat \eta_i+S_i^z\hat \zeta_i\label{SI}\\
\vec S_j&=&S_j^x\hat \xi_j+S_j^y\hat \eta_j+S_j^z\hat \zeta_j\label{SJ}
\end{eqnarray}
We have (see Fig. \ref{local})
\begin{eqnarray}
\hat \xi_j&=&\cos \theta_{ij}\hat \zeta_i+\sin \theta_{ij}\hat \xi_i\nonumber\\
\hat \zeta_j&=&-\sin \theta_{ij}\hat \zeta_i+\cos \theta_{ij}\hat \xi_i\nonumber\\
\hat \eta_j&=&\hat \eta_i\nonumber
\end{eqnarray}
where $\cos \theta_{ij}=\cos (\theta_i-\theta_j)$ is the angle between two
spins $i$ and $j$.
Replacing these into Eq. (\ref{SJ}) to express $\vec S_j$ in the $(\hat \xi_i,\hat \eta_i,\hat \zeta_i)$ coordinates, then calculating
$\vec S_i \cdot \vec S_j$, we obtain the following exchange Hamiltonian from (\ref{eqn:hamil1}):
\begin{eqnarray}
\mathcal H_e &=& - \sum_{<i,j>}
J_{i,j}\Bigg\{\frac{1}{4}\left(\cos\theta_{ij} -1\right)
\left(S^+_iS^+_j +S^-_iS^-_j\right)\nonumber\\
&+& \frac{1}{4}\left(\cos\theta_{ij} +1\right) \left(S^+_iS^-_j
+S^-_iS^+_j\right)\nonumber\\
&+&\frac{1}{2}\sin\theta_{ij}\left(S^+_i +S^-_i\right)S^z_j
-\frac{1}{2}\sin\theta_{ij}S^z_i\left(S^+_j
+S^-_j\right)\nonumber\\
&+&\cos\theta_{ij}S^z_iS^z_j\Bigg\}
\label{eq:HGH2}
\end{eqnarray}
At this stage, let us mention that according to the theorem of Mermin and Wagner \cite{Mermin}
 continuous isotropic spin models such as XY and Heisenberg spins
do not have long-range ordering at finite temperatures in two dimensions. Since we are dealing with the Heisenberg model in a thin film, it is useful to add an anisotropic interaction  to create a long-range ordering and a phase transition at finite temperatures.
We choose the following anisotropic interaction along the in-plane local spin-quantization axes $z$ of $\mathbf S_i$
and $\mathbf S_j$:
\begin{equation}
\mathcal H_a= -\sum_{<i,j>} I_{i,j}S^z_iS^z_j\cos\theta_{ij}
\end{equation}
where $I_{i,j}(>0)$ is supposed to be positive, small compared to $J_1$, and limited to NN on the $c$-axis.
The full Hamiltonian is thus
$\mathcal H=\mathcal H_e+\mathcal H_a$.

\subsection{General formulation for non collinear magnets}
We define the following two double-time Green's functions in the real space:
\begin{eqnarray}
G_{i,j}(t,t')&=&<<S_i^+(t);S_{j}^-(t')>>\nonumber\\
&=&-i\theta (t-t')
<\left[S_i^+(t),S_{j}^-(t')\right]> \label{green59a}\\
F_{i,j}(t,t')&=&<<S_i^-(t);S_{j}^-(t')>>\nonumber\\
&=&-i\theta (t-t')
<\left[S_i^-(t),S_{j}^-(t')\right]>\label{green60}
\end{eqnarray}
We need these two functions because the equation of motion of the first function generates functions of
the second type, and vice-versa. These equations of motion are
\begin{eqnarray}
i\hbar\frac {d}{dt}G_{i,j}\left( t,t'\right) &=& \left<\left[ S^+_i
\left( t\right) , S^-_j \left( t'\right)\right]\right>\delta\left(
t-t'\right) \nonumber\\
&-& \left<\left< \left[\mathcal H, S^+_i\left( t\right)\right] ;
S^-_j \left( t'\right) \right>\right>,
\label{eq:HGEoMG}\\
i\hbar\frac {d}{dt}F_{i,j}\left( t,t'\right) &=& \left<\left[ S^-_i
\left( t\right) , S^-_j \left( t'\right)\right]\right>\delta\left(
t-t'\right)\nonumber \\
&-& \left<\left< \left[\mathcal H, S^-_i\left( t\right)\right] ;
S^-_j \left( t'\right) \right>\right>, \label{eq:HGEoMF}
\end{eqnarray}
Expanding the commutators, and using the Tyablikov decoupling scheme for
higher-order functions, for example $<<S_{i'}^zS_i^+(t);S_{j}^-(t')>>\simeq <S_{i'}^z><<S_i^+(t);S_{j}^-(t')>>$ etc., we
obtain the following general equations for non collinear magnets:
 \begin{eqnarray}
 i\hbar \frac{dG_{i,j}(t,t')}{dt}&=&2<S_i^z>\delta_{i,j} \delta (t-t')\nonumber\\
 &-&\sum_{i'}J_{i,i'}[<S_i^z>(\cos \theta_{i,i'}-1)\times \nonumber\\
 &\times& F_{i',j}(t,t')\nonumber\\
&+&<S_i^z>(\cos \theta_{i,i'}+1)G_{i',j}(t,t')\nonumber\\
&-&2<S_{i'}^z>\cos \theta_{i,i'}G_{i,j}(t,t')]\nonumber\\
&+&2\sum_{i'}I_{i,i'}<S_{i'}^z>\cos \theta_{i,i'}G_{i,j}(t,t')\nonumber\\
&&\label{GFG0}\\
i\hbar \frac{dF_{i,j}(t,t')}{dt}&=&\sum_{i'}J_{i,i'}[<S_i^z>(\cos \theta_{i,i'}-1)\times\nonumber\\
&\times& G_{i',j}(t,t')\nonumber\\
&+&<S_i^z>(\cos \theta_{i,i'}+1)F_{i',j}(t,t')\nonumber\\
&-&2<S_{i'}^z>\cos \theta_{i,i'}F_{i,j}(t,t')]\nonumber\\
&-&2\sum_{i'}I_{i,i'}<S_{i'}^z>\cos \theta_{i,i'}F_{i,j}(t,t')\nonumber\\
&&\label{GFF0}
\end{eqnarray}

\subsection{BCC helimagnetic films}

In the case of a BCC thin film with a (001) surface,  the above equations yield a closed system of
coupled equations within the Tyablikov decoupling scheme.
For clarity, we  separate the sums on NN interactions and NNN interactions as follows:
 \begin{eqnarray}
 i\hbar \frac{dG_{i,j}(t,t')}{dt}&=&2<S_i^z>\delta_{i,j} \delta (t-t')\nonumber\\
 &-&\sum_{k'\in NN}J_{i,k'}[<S_i^z>(\cos \theta_{i,k'}-1) \times \nonumber\\
 &\times& F_{k',j}(t,t')\nonumber\\
 &+&<S_i^z>(\cos \theta_{i,k'}+1) G_{k',j}(t,t')\nonumber\\
 &-&2<S_{k'}^z>\cos \theta_{i,k'}G_{i,j}(t,t')]\nonumber\\
 &+&2\sum_{k'\in NN}I_{i,k'}<S_{k'}^z>\cos \theta_{i,k'}G_{i,j}(t,t')\nonumber\\
&-&\sum_{i'\in NNN}J_{i,i'}[<S_i^z>(\cos \theta_{i,i'}-1) \times \nonumber\\
&\times& F_{i',j}(t,t')\nonumber\\
&+&<S_i^z>(\cos \theta_{i,i'}+1) G_{i',j}(t,t')\nonumber\\
&-&2<S_{i'}^z>\cos \theta_{i,i'}G_{i,j}(t,t')]\label{GFG2}\\
i\hbar \frac{dF_{k,j}(t,t')}{dt}&=&\sum_{i'\in NN}J_{k,i'}[<S_k^z>(\cos \theta_{k,i'}-1)\times\nonumber\\
&\times&  G_{i',j}(t,t')\nonumber\\
&+&<S_k^z>(\cos \theta_{k,i'}+1) F_{i',j}(t,t')\nonumber\\
&-&2<S_{i'}^z>\cos \theta_{k,i'}F_{k,j}(t,t')]\nonumber\\
&-&2\sum_{i'\in NN}I_{k,i'}<S_{i'}^z>\cos \theta_{k,i'}F_{k,j}(t,t')\nonumber\\
&+&\sum_{k'\in NNN}J_{k,k'}[<S_k^z>(\cos \theta_{k,k'}-1) \times \nonumber\\
&\times& G_{k',j}(t,t')\nonumber\\
&+&<S_k^z>(\cos \theta_{k,k'}+1) F_{k',j}(t,t')\nonumber\\
&-&2<S_{k'}^z>\cos \theta_{k,k'}F_{k,j}(t,t')]\label{GFF2}
\end{eqnarray}
For simplicity, except otherwise stated, all NN interactions $(J_{k,k'}, I_{k,k'})$ are taken equal to $(J_1,I_1)$ and all NNN interactions
are taken equal to $J_2$ in the following.  Furthermore, let us define the film coordinates which are used
below: the $c$-axis is called $z$-axis, planes parallel to the film surface are called $xy$-planes and the Cartesian components
of the spin position $\mathbf R_i$ are denoted by $(\ell_i,m_i,n_i)$.

% \begin{eqnarray}
% i\hbar \frac{dG_{i,j}(t,t')}{dt}&=&2<S_i^z>\delta_{i,j} \delta (t-t')\nonumber\\
% &&-\sum_{k'}J_{i,k'}[<S_i^z>(\cos \theta_{i,k'}-1) \times \nonumber\\
% &&\times F_{k',j}(t,t')\nonumber\\
% &&-2<S_{k'}^z>\cos \theta_{i,k'}G_{i,j}(t,t')]\nonumber\\
%&& -\sum_{i'}J_{i,i'}[<S_i^z>(\cos \theta_{i,i'}+1) \times \nonumber\\
%&&\times G_{i',j}(t,t')\nonumber\\
%&&-2<S_{i'}^z>\cos \theta_{i,i'}G_{i,j}(t,t')]\nonumber\\
%&&+I_i(2<S_i^z>-1)G_{i,j}(t,t')\label{GFG}\\
%i\hbar \frac{dF_{k,j}(t,t')}{dt}&=&\sum_{i'}J_{k,i'}[<S_k^z>(\cos \theta_{k,i'}-1)\times\nonumber\\
%&&\times G_{i',j}(t,t')\nonumber\\
%&&-2<S_{i'}^z>\cos \theta_{k,i'}F_{k,j}(t,t')]\nonumber\\
%&&+\sum_{k'}J_{k,k'}[<S_k^z>(\cos \theta_{k,k'}+1)\times\nonumber\\
%&&\times  F_{k',j}(t,t')\nonumber\\
%&&-2<S_{k'}^z>\cos \theta_{k,k'}F_{k,j}(t,t')]\nonumber\\
%&&-I_k(2<S_k^z>+1)F_{k,j}(t,t')\label{GFF}
%\end{eqnarray}
We now introduce the following in-plane Fourier
transforms:

\begin{eqnarray}
G_{i, j}\left( t, t'\right) &=& \frac {1}{\Delta}\int\int_{BZ} d\mathbf
k_{xy}\frac{1}{2\pi}\int^{+\infty}_{-\infty}d\omega e^{-i\omega
\left(t-t'\right)}\nonumber\\
&&\hspace{0.7cm}\times g_{n_i,n_j}\left(\omega , \mathbf k_{xy}\right)
e^{i\mathbf k_{xy}\cdot \left(\mathbf R_i-\mathbf
R_j\right)},\label{eq:HGFourG}\\
F_{k, j}\left( t, t'\right) &=& \frac {1}{\Delta}\int\int_{BZ} d\mathbf
k_{xy}\frac{1}{2\pi}\int^{+\infty}_{-\infty}d\omega e^{-i\omega
\left(t-t'\right)}\nonumber\\
&&\hspace{0.7cm}\times f_{n_k,n_j}\left(\omega , \mathbf k_{xy}\right)
e^{i\mathbf k_{xy}\cdot \left(\mathbf R_k-\mathbf
R_j\right)},\label{eq:HGFourF}
\end{eqnarray}
where $\omega$ is the spin-wave frequency, $\mathbf k_{xy}$
denotes the wave-vector parallel to $xy$ planes and $\mathbf R_i$ is
the position of the spin at the site $i$. $n_i$, $n_j$ and $n_k$ are
respectively the $z$-component indices of the layers where the sites $\mathbf R_i$,  $\mathbf R_j$ and $\mathbf R_k$
belong to. The integral over $\mathbf k_{xy}$ is performed in the
first Brillouin zone ($BZ$) whose surface is $\Delta$ in the $xy$
reciprocal plane.  For convenience, we denote $n_i=1$ for all sites on the surface layer, $n_i=2$ for all sites of the second layer and so on.

Note that for a three-dimensional case,  making a 3D Fourier transformation  of Eqs. (\ref{GFG2})-(\ref{GFF2})
we obtain the spin-wave dispersion relation in the absence of anisotropy:
\begin{equation}
\hbar\omega=\pm \sqrt{A^2-B^2}
\end{equation}
where
\begin{eqnarray}
A&=& J_1 \left< S^z\right>[\cos\theta+1]Z\gamma+2Z J_1\left< S^z\right> \cos \theta\nonumber\\
&&+J_2 \left< S^z\right> [\cos (2\theta)+1]Z_c\cos (k_za) \nonumber\\
&&+2Z_cJ_2 \left< S^z\right> \cos (2\theta)\nonumber\\
B&=& J_1 \left< S^z\right> (\cos\theta -1)Z\gamma\nonumber\\
&&+J_2 \left< S^z\right>[\cos(2\theta) -1]Z_c\cos (k_za) \nonumber
\end{eqnarray}
where $Z=8$ (NN number), $Z_c=2$ (NNN number on the $c$-axis), $\gamma=\cos (k_xa/2)\cos (k_ya/2)\cos (k_za/2)$ ($a$: lattice constant).
We see that $\hbar\omega$ is zero when $A=\pm B$, namely at $k_x=k_y=k_z=0$ ($\gamma=1$) and at $k_z=2\theta$ along the helical axis.
The case of ferromagnets (antiferromagnets) with NN interaction only is recovered by putting $\cos \theta=1$ $(-1)$ \cite{Diep1979}.

Let us return to the film case. We make the in-plane Fourier transformation Eqs. (\ref{eq:HGFourG})-(\ref{eq:HGFourF})
for  Eqs. (\ref{GFG2})-(\ref{GFF2}). We obtain the following matrix equation
\begin{equation}
\mathbf M \left( \omega \right) \mathbf h = \mathbf u,
\label{eq:HGMatrix}
\end{equation}
where $\mathbf M\left(\omega\right)$ is a square matrix of dimension
$\left(2N_z \times 2N_z\right)$, $\mathbf h$ and $\mathbf u$ are
the column matrices which are defined as follows
\begin{equation}
\mathbf h = \left(%
\begin{array}{c}
  g_{1,n'} \\
  f_{1,n'} \\
  \vdots \\
  g_{n,n'} \\
  f_{n,n'} \\
    \vdots \\
  g_{N_z,n'} \\
  f_{N_z,n'} \\
\end{array}%
\right) , \mathbf u =\left(%
\begin{array}{c}
  2 \left< S^z_1\right>\delta_{1,n'}\\
  0 \\
  \vdots \\
  2 \left< S^z_{N_z}\right>\delta_{N_z,n'}\\
  0 \\
\end{array}%
\right) , \label{eq:HGMatrixgu}
\end{equation}
where,  taking $\hbar=1$ hereafter,
\begin{widetext}
\begin{equation}
\mathbf M\left(\omega\right) = \left(%
\begin{array}{cccccccccccc}
  \omega+A_1&0    & B^+_1    & C^+_1& D_1^+& E_1^+& 0&0&0&0&0&0\\
   0    & \omega-A_1  & -C^+_1 & -B^+_1 &-E_1^+&-D_1^+&0&0&0&0&0&0\\
   \cdots & \cdots & \cdots &\cdots&\cdots&\cdots&\cdots&\cdots&\cdots&\cdots&\cdots&\cdots\\
 \cdots&D_n^-&E_n^-&B^-_{n}&C^-_{n}&\omega+A_{n}&0&B^+_{n}&C^+_{n}&D_n^+&E_n^+&\cdots\\
 \cdots&-E_n^-&-D_n^-&-C^-_{n}&-B^-_{n}&0&\omega-A_{n}&-C^+_{n}&-B^+_{n}&-E_n^+&-D_n^+&\cdots\\
         \cdots  & \cdots & \cdots & \cdots &\cdots&\cdots&\cdots&\cdots&\cdots&\cdots&\cdots&\cdots \\
  0& 0&0&0& 0& 0& D^-_{N_z}& E^-_{N_z}  & B^-_{N_z}   & C^-_{N_z}   &\omega + A_{N_z}&0\\
  0&0&0&0&0&0&-E^-_{N_z}& -D^-_{N_z} & -C^-_{N_z}  & -B^-_{N_z}&0  & \omega-A_{N_z}\\
\end{array}%
\right) \label{eq:HGMatrixM}
\end{equation}
\end{widetext}
where
\begin{eqnarray}
A_{n} &=& - 8J_1(1+d) \Big[\left< S^z_{n+1}\right>
\cos\theta_{n,n+1}\nonumber\\
&&+\left< S^z_{n-1}\right>
\cos\theta_{n,n-1}\Big]\nonumber\\
&-&2J_2 \Big[\left< S^z_{n+2}\right>
\cos\theta_{n,n+2}\nonumber\\
&&+\left< S^z_{n-2}\right>
\cos\theta_{n,n-2}\Big]\nonumber
\end{eqnarray}
where $n=1,2,...,N_z$, $d=I_1/J_1$, and
\begin{eqnarray}
B_n^\pm &=& 4J_1 \left< S^z_{n}\right>(\cos\theta_{n,n\pm 1}+1)\gamma \nonumber\\
C_n^\pm &=& 4J_1 \left< S^z_{n}\right>(\cos\theta_{n,n\pm 1}-1)\gamma \nonumber\\
E_n^\pm &=& J_2 \left< S^z_{n}\right>(\cos\theta_{n,n\pm 2}-1)\nonumber\\
D_n^\pm &=& J_2 \left< S^z_{n}\right>(\cos\theta_{n,n\pm 2}+1) \nonumber
\end{eqnarray}
In the above expressions, $\theta_{n,n\pm
1}$ the angle between a spin in the layer $n$ and its NN spins in layers $n\pm 1$ etc. and
$\gamma = \cos \left( \frac{k_x a}{2} \right)\cos \left( \frac{k_y a}{2} \right).$

Solving det$|\mathbf M|=0$, we obtain the spin-wave spectrum
$\omega$ of the present system: for each value ($k_x,k_y)$, there are 2$N_z$ eigen-values of $\omega$ corresponding to two opposite spin precessions as in antiferromagnets (the dimension of det$|\mathbf M|$ is $2N_z\times 2N_z$).  Note that the above equation depends on the values of $<S_n^z>$ ($n=1,...,N_z$).
Even at temperature $T=0$, these $z$-components are not equal to $1/2$ because we are dealing with an antiferromagnetic
system where fluctuations at $T=0$ give rise to the so-called zero-point spin contraction \cite{DiepTM}. Worse, in our system
with the existence of the film surfaces, the spin contractions are not spatially uniform as will be seen below.
So the solution of det$|\mathbf M|=0$ should be found by iteration.  This will be explicitly shown hereafter.

The solution for $g_{n,n}$ is given by
\begin{equation}
g_{n,n}(\omega) = \frac{\left|\mathbf M\right|_{2n-1}}{\left|\mathbf M\right|},
\end{equation}
where $\left|\mathbf M\right|_{2n-1}$ is the determinant made by
replacing the $2n-1$-th column of $\left|\mathbf M\right|$ by
$\mathbf u$ given by Eq. (\ref{eq:HGMatrixgu}) [note that $g_{n,n}$ occupies the $(2n-1)$-th line of the matrix $\mathbf h$]. Writing now
\begin{equation}
\left|\mathbf M\right| = \prod_i \left[\omega -
\omega_i\left(\mathbf k_{xy}\right)\right],
\end{equation}
we see that $\omega_i\left(\mathbf k_{xy}\right) ,\ i = 1,\cdots
,\ 2N_z$, are poles of  $g_{n,n}$.
$\omega_i\left(\mathbf k_{xy}\right)$ can be obtained by solving
$\left|\mathbf M\right|=0$. In this case, $g_{n,n}$ can be
expressed as
\begin{equation}
g_{n, n}(\omega) = \sum_i\frac {D_{2n-1}\left(\omega_i\left(\mathbf
k_{xy}\right)\right)}{\left[ \omega - \omega_i\left(\mathbf
k_{xy}\right)\right]}, \label{eq:HGGnn}
\end{equation}
where $D_{2n-1}\left(\omega_i\left(\mathbf k_{xy}\right)\right)$ is
\begin{equation}
D_{2n-1}\left(\omega_i\left(\mathbf k_{xy}\right)\right) = \frac{\left|
\mathbf M\right|_{2n-1} \left(\omega_i\left(\mathbf
k_{xy}\right)\right)}{\prod_{j\neq i}\left[\omega_j\left(\mathbf
k_{xy}\right)-\omega_i\left(\mathbf k_{xy}\right)\right]}.
\end{equation}

Next, using the spectral theorem which relates the correlation
function \(\langle S^-_i S^+_j\rangle \) to the Green's function \cite{zu}, we have
\begin{eqnarray}
\left< S^-_i S^+_j\right> &=& \lim_{\varepsilon\rightarrow 0}
\frac{1}{\Delta}\int\int d\mathbf k_{xy}
\int^{+\infty}_{-\infty}\frac{i}{2\pi}\big( g_{n, n'}\left(\omega
+ i\varepsilon\right)\nonumber\\
&-& g_{n, n'}\left(\omega - i\varepsilon\right)\big)
\frac{d\omega}{e^{\beta\omega} - 1}e^{i\mathbf
k_{xy}\cdot\left(\mathbf R_i -\mathbf R_j\right)},
\end{eqnarray}
where $\epsilon$ is an  infinitesimal positive constant and
$\beta=(k_BT)^{-1}$, $k_B$ being the Boltzmann constant.

%\begin{figure}[htb]
%\centering
%\includegraphics[width=8cm]{helispec.eps}  % .eps
%\caption{\label{ffighelispec} Spin-wave spectrum versus the wave-vector
%$k_y$ in the simple cubic lattice with a helical structure in the $y$ axis,
%in the case $\theta=\pi/3$ (namely $|J_2|/J_1=2$), and $k_x=k_z=0$.}
%\end{figure}

Using the Green's function presented above, we can calculate
self-consistently various physical quantities as functions of
temperature $T$.  The magnetization $\langle S_{n}^z\rangle$ of the $n$-th layer is given by
\begin{eqnarray}
\langle S_{n}^z\rangle&=&\frac{1}{2}-\left< S^-_{n} S^+_{n}\right>\nonumber\\
&=&\frac{1}{2}-
   \lim_{\epsilon\to 0}\frac{1}{\Delta}
   \int
   \int d{\bf k_{xy}}
   \int\limits_{-\infty}^{+\infty}\frac{i}{2\pi}
   [ g_{n,n}(\omega+i\epsilon)\nonumber\\
   &&-g_{n,n}(\omega-i\epsilon)]
\frac{d\omega}{\mbox{e}^{\beta \omega}-1}\label{lm1}
\end{eqnarray}
Replacing Eq. (\ref{eq:HGGnn}) in Eq. (\ref{lm1}) and making use of the following identity

\begin{equation}\label{id}
\frac {1}{x-i\eta} - \frac {1}{x+i\eta}=2\pi i\delta (x)
\end{equation}
we obtain
\begin{equation}\label{lm2}
\langle S_{n}^z\rangle=\frac{1}{2}-
   \frac{1}{\Delta}
   \int
   \int dk_xdk_y
   \sum_{i=1}^{2N_z}\frac{D_{2n-1}(\omega_i)}
   {\mbox{e}^{\beta \omega_i}-1}
\end{equation}
where $n=1,...,N_z$.
As $<S_{n}^z>$ depends on the magnetizations of the neighboring layers via $\omega_i (i=1,...,2N_z)$,
we should solve by iteration the equations
(\ref{lm2}) written for all layers, namely for  $n=1,...,N_z$, to obtain the magnetizations of layers 1, 2, 3, ..., $N_z$
at a given temperature $T$. Note that by symmetry, $<S_1^z>=<S_{N_z}^z>$, $<S_{2}^z>=<S_{N_z-1}^z>$, $<S_3^z>=<S_{N_z-2}^z>$, and so on.
Thus, only $N_z/2$ self-consistent layer magnetizations are to be calculated.

The value of the spin in the layer $n$ at $T=0$ is calculated by

\begin{equation}\label{surf38}
\langle S_{n}^z\rangle(T=0)=\frac{1}{2}+
   \frac{1}{\Delta} \int \int dk_xdk_y
   \sum_{i=1}^{N_z}D_{2n-1}(\omega_i)
\end{equation}
where the sum is performed over $N_z$ negative values of  $\omega_i$ (for positive values the Bose-Einstein factor is equal to 0 at $T=0$).

The transition temperature $T_c$ can be calculated in a self-consistent manner by iteration, letting all  $<S_{n}^z>$  tend to zero, namely $\omega_i\rightarrow 0$. Expanding $\mbox{e}^{\beta \omega_i}-1\rightarrow  \beta_c \omega_i$ on the right-hand side of Eq. (\ref{lm2}) where $\beta_c=(k_BT_c)^{-1}$, we have by putting $\langle S_{n}^z\rangle=0$ on the left-hand side,
\begin{equation}\label{tcc}
\beta_c=2\frac{1}{\Delta}\int \int dk_xdk_y
   \sum_{i=1}^{2N_z}\frac{D_{2n-1}(\omega_i)}{\omega_i}
\end{equation}
There are $N_z$ such equations using Eq. (\ref{lm2}) with $n=1,...,N_z$.  Since the layer magnetizations tend to zero at the transition temperature from different values, it is obvious that we have to look for a convergence of the solutions of the equations Eq. (\ref{tcc}) to a single value of $T_c$. The method to do this will be shown below.

\section{Results from the Green's function method}
Let us take $J_1=1$, namely ferromagnetic interaction between NN.  We consider the helimagnetic case where the NNN interaction $J_2$ is negative and
$|J_2|>J_1$. The non uniform GS spin configuration across the film has been determined in section \ref{GSSC} for each value of $p=J_2/J_1$.   Using the values of $\theta_{n,n\pm1}$ and $\theta_{n,n\pm2}$ to calculate the matrix elements of $\left|\mathbf M\right|$, then solving det$\left|\mathbf M\right|=0$ we find the eigenvalues $\omega_i(i=1,...,2N_z)$ for each $\mathbf k_{xy}$ with a input set of $\langle S_{n}^z\rangle (n=1,...,N_z)$ at a given $T$. Using
Eq. (\ref{lm2}) for $n=1,...,N_z$ we calculate the output  $\langle S_{n}^z\rangle (n=1,...,N_z)$. Using this output set as input, we calculate again
 $\langle S_{n}^z\rangle (n=1,...,N_z)$ until the input and output are identical within a desired precision $P$.  Numerically, we use a Brillouin zone of $100^2$ wave-vector values, and we use the obtained values $\langle S_{n}^z\rangle$ at  a given $T$ as input for a neighboring $T$. At low $T$ and up to $\sim \frac{4}{5} T_c$, only a few iterations suffice to get $P\leq 1\%$. Near $T_c$, several dozens of iteration are needed to get convergence.  We show below our results.

\subsection{Spectrum}

\begin{figure}[htb]
\centering
\includegraphics[width=7cm,angle=0]{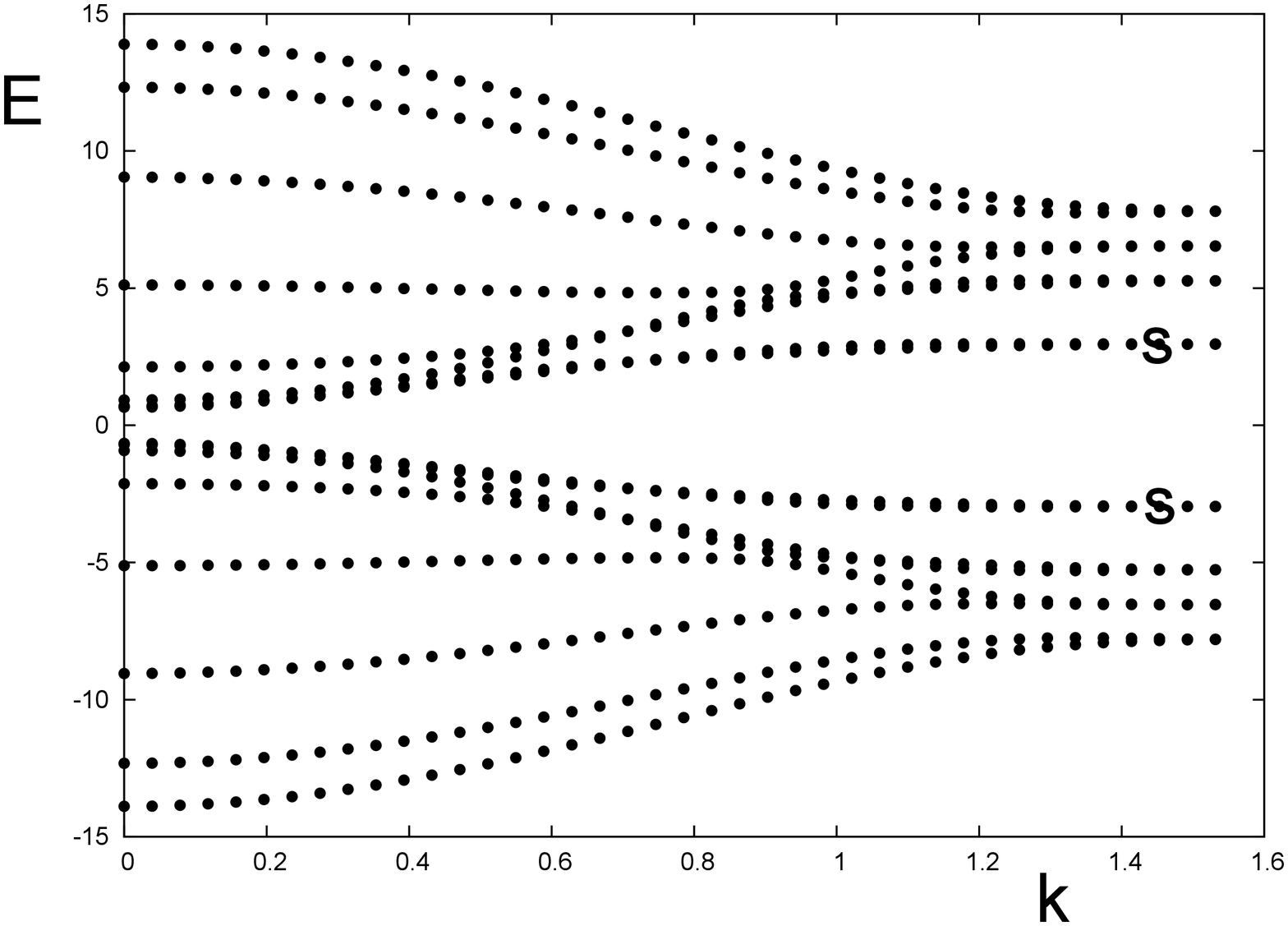}  % .eps
\includegraphics[width=7cm,angle=0]{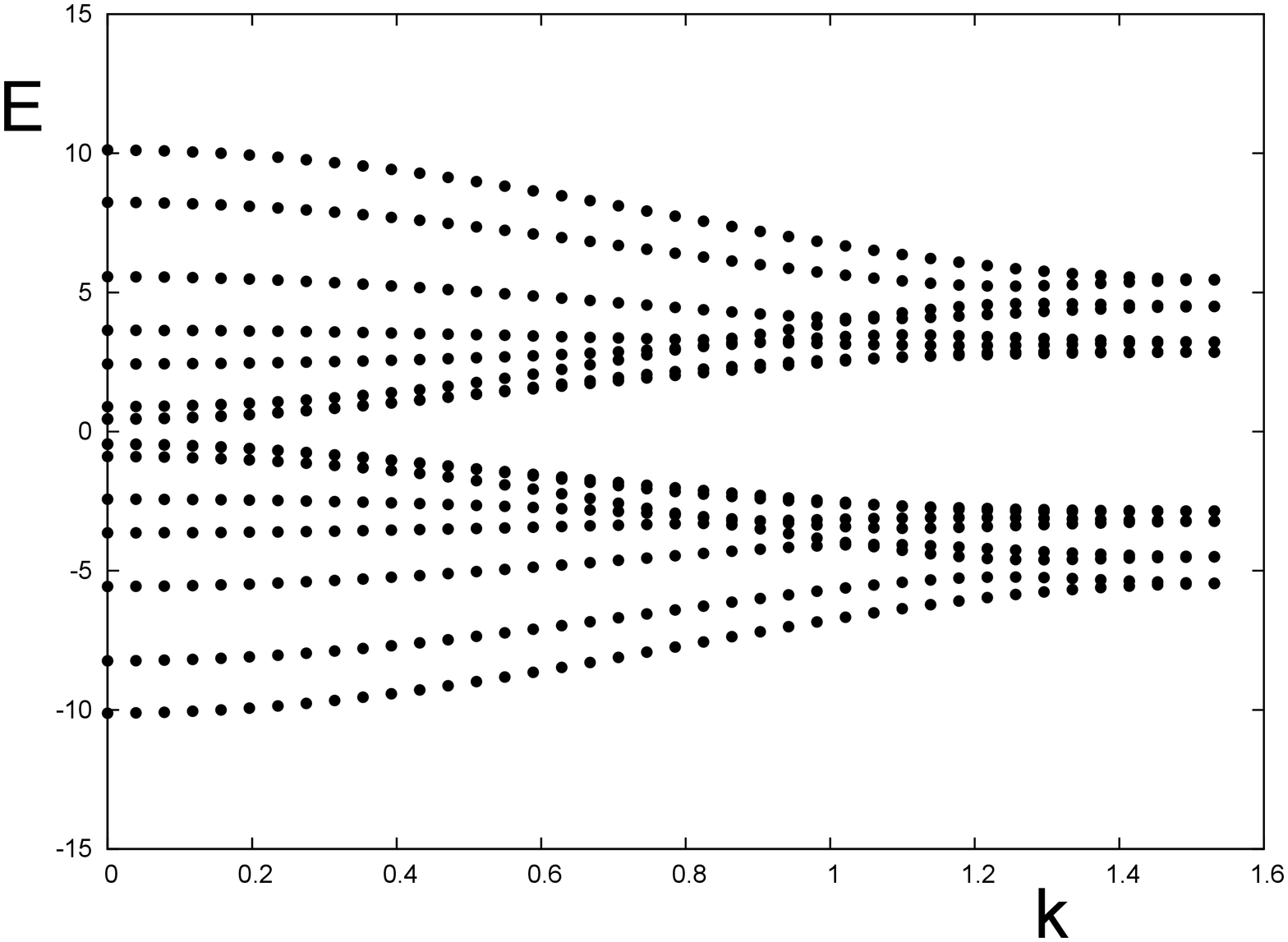}  % .eps
\caption{ Spectrum $E=\hbar \omega$ versus $k\equiv k_x=k_y$ for $J_2/J_1=-1.4$ at $T=0.1$ (top) and $T=1.02$ (bottom) for $N_z=8$ and $d=0.1$. The surface branches are indicated by $s$.}\label{sweta}
\end{figure}

We calculated the spin-wave spectrum as described above for each a given $J_2/J_1$. The spin-wave spectrum depends on the temperature via the temperature-dependence of layer magnetizations. Let us show in Fig. \ref{sweta} the spin-wave frequency $\omega$ versus $k_x=k_y$ in the case on an 8-layer film where $J_2/J_1=-1.4$  at two temperatures $T=0.1$ and $T=1.02$ (in units of $J_1/k_B=1$).   There are 8 positive and 8 negative modes corresponding two opposite spin precessions. Note that there are two degenerate acoustic surface branches lying at low energy on each side. This degeneracy comes from the two symmetrical surfaces of the film. These surface modes propagate parallel to the film surface but are damped from the surface inward.   As $T$ increases, layer magnetizations decrease (see below), reducing therefore the spin-wave energy as seen in Fig. \ref{sweta} (bottom).

%\subsection{Effect of $\varphi=J_2/J_1$}

\subsection{Spin contraction at $T=0$ and transition temperature}
\begin{figure}[htb]
\centering
\includegraphics[width=7cm,angle=0]{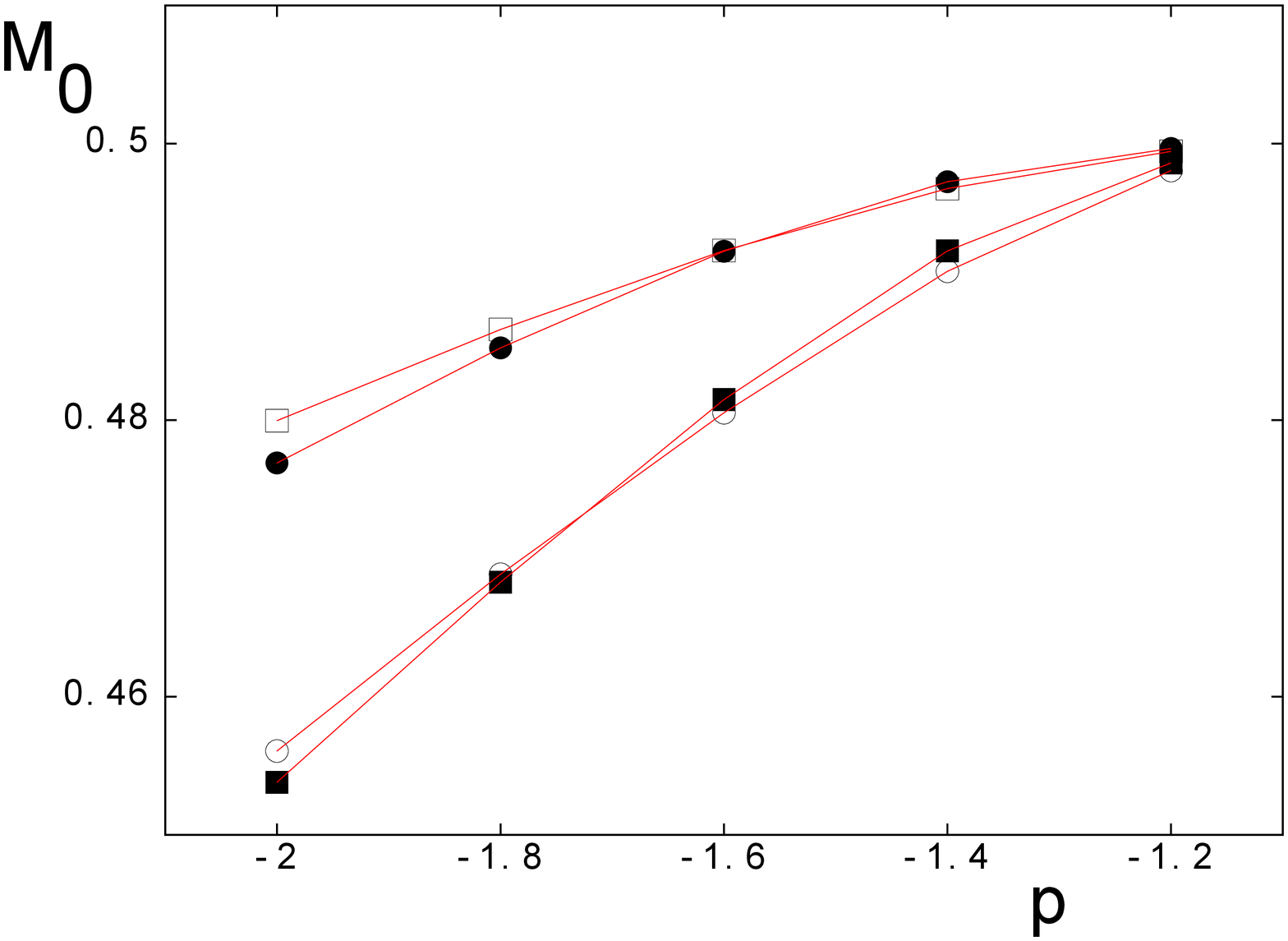}  % .eps
\caption{(Color online) Spin lengths of the first four layers at $T=0$ for several values of $p=J_2/J_1$ with $d=0.1$, $N_z=8$.
Black circles,  void circles, black squares and void squares are for first, second, third and fourth layers, respectively. See text for comments. }\label{spin0}
\end{figure}

It is known that in antiferromagnets, quantum fluctuations give rise to a contraction of the spin length at zero temperature \cite{DiepTM}.  We will see here that a spin under a stronger antiferromagnetic interaction has a stronger zero-point spin contraction. The spins near the surface serve for such a test. In the case of the film considered above, spins in the first and in the second layers have only one antiferromagnetic NNN while interior spins have two NNN, so the contraction at a given $J_2/J_1$ is expected to be stronger for interior spins. This is verified with the results shown in Fig. \ref{spin0}.   When $|J_2|/J_1$ increases, namely the antiferromagnetic interaction becomes stronger, we observe  stronger contractions. Note that the contraction tends to zero when the spin configuration becomes ferromagnetic, namely $J_2$ tends to -1.

\subsection{Layer magnetizations}

\begin{figure}[htb]
\centering
\includegraphics[width=7cm,angle=0]{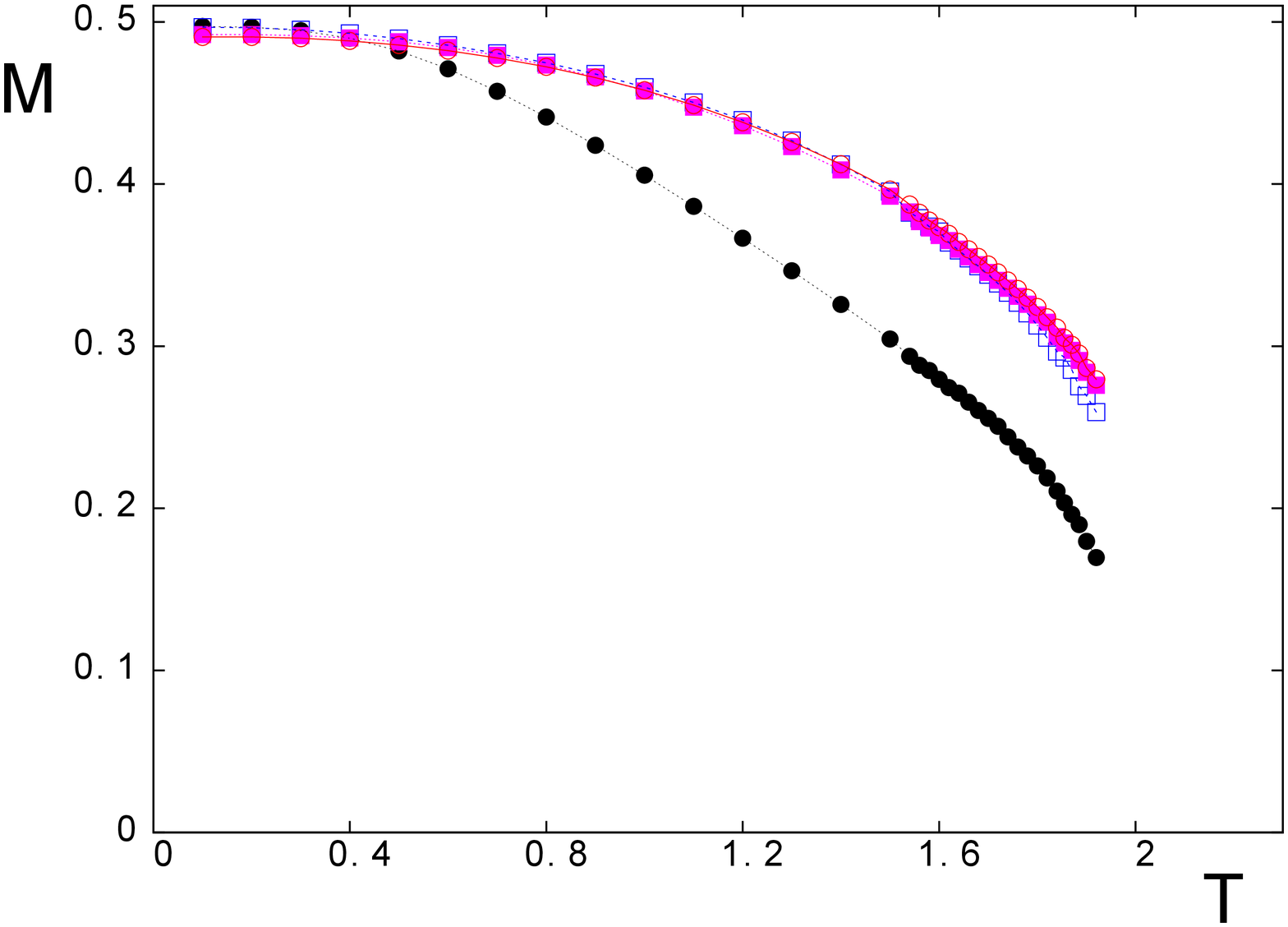}  % .eps
\includegraphics[width=7cm,angle=0]{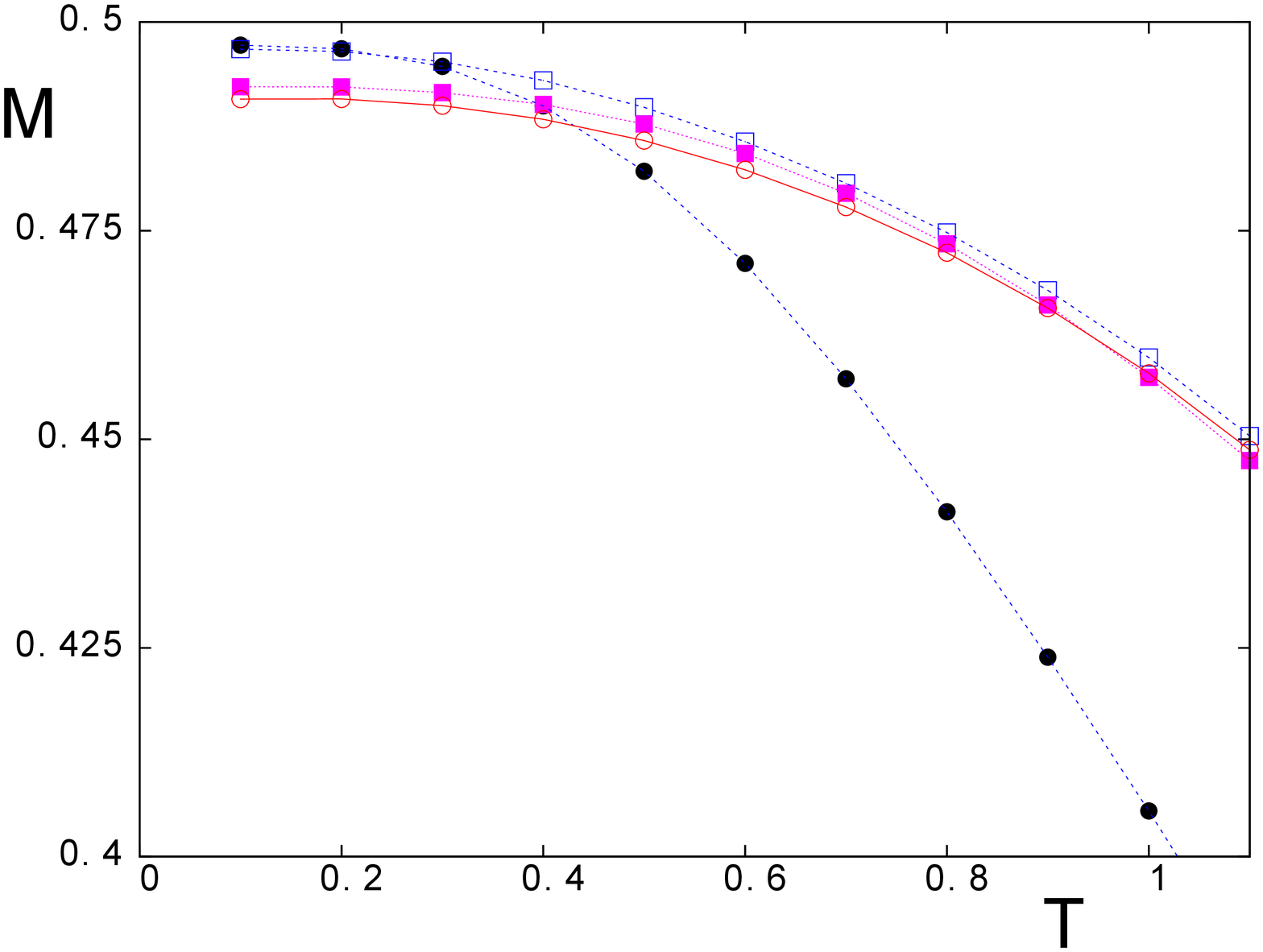}  % .eps
\caption{(Color online) Layer magnetizations as functions of $T$ for $J_2/J_1=-1.4$ with $d=0.1$, $N_z=8$ (top). Zoom of the region at low $T$ to show crossover (bottom). Black circles, blue void squares, magenta squares and red void circles are for first, second, third and fourth layers, respectively.  See text.}\label{magnet14}
\end{figure}

\begin{figure}[htb]
\centering
\includegraphics[width=7cm,angle=0]{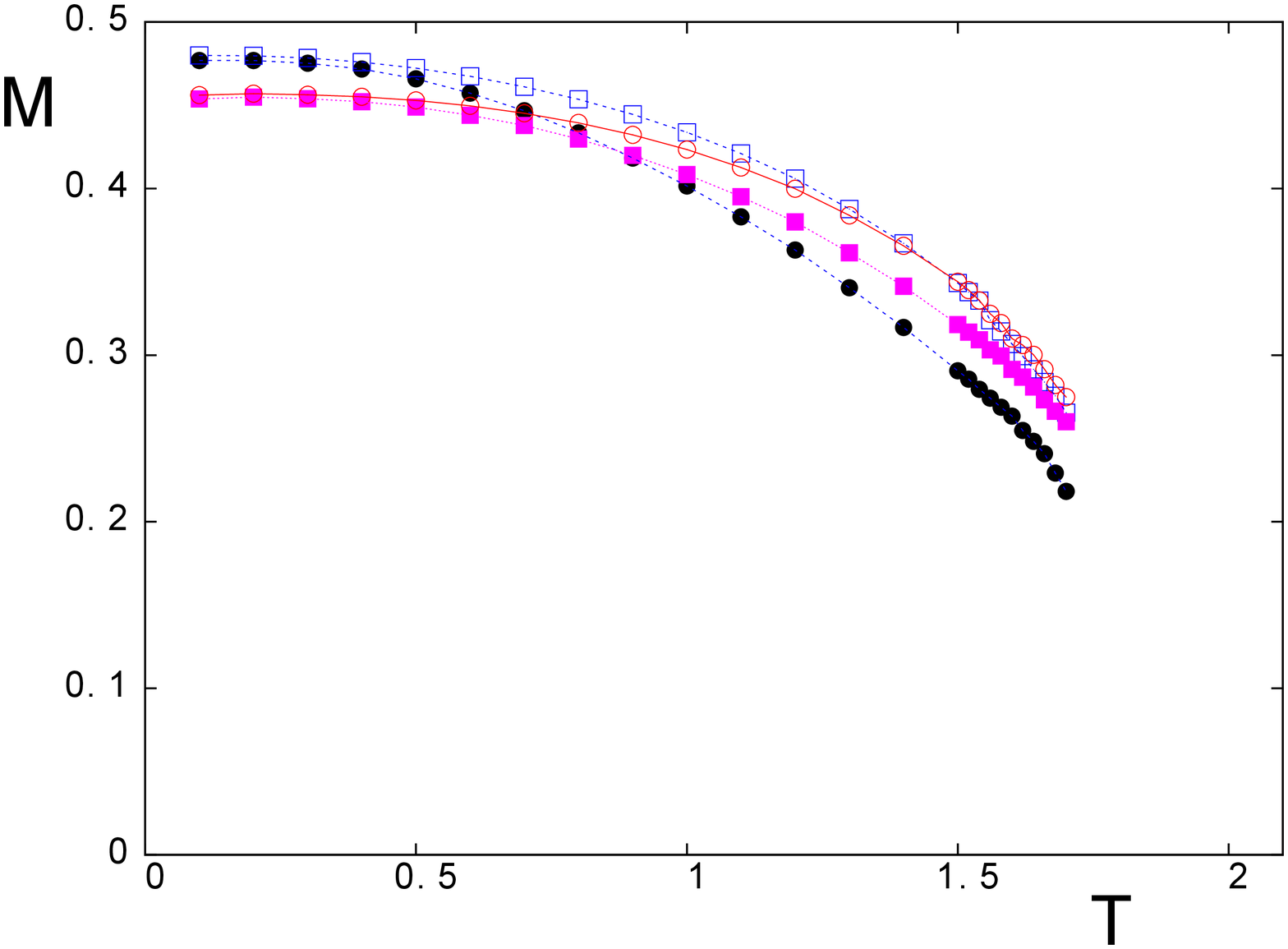}  % .eps
\includegraphics[width=7cm,angle=0]{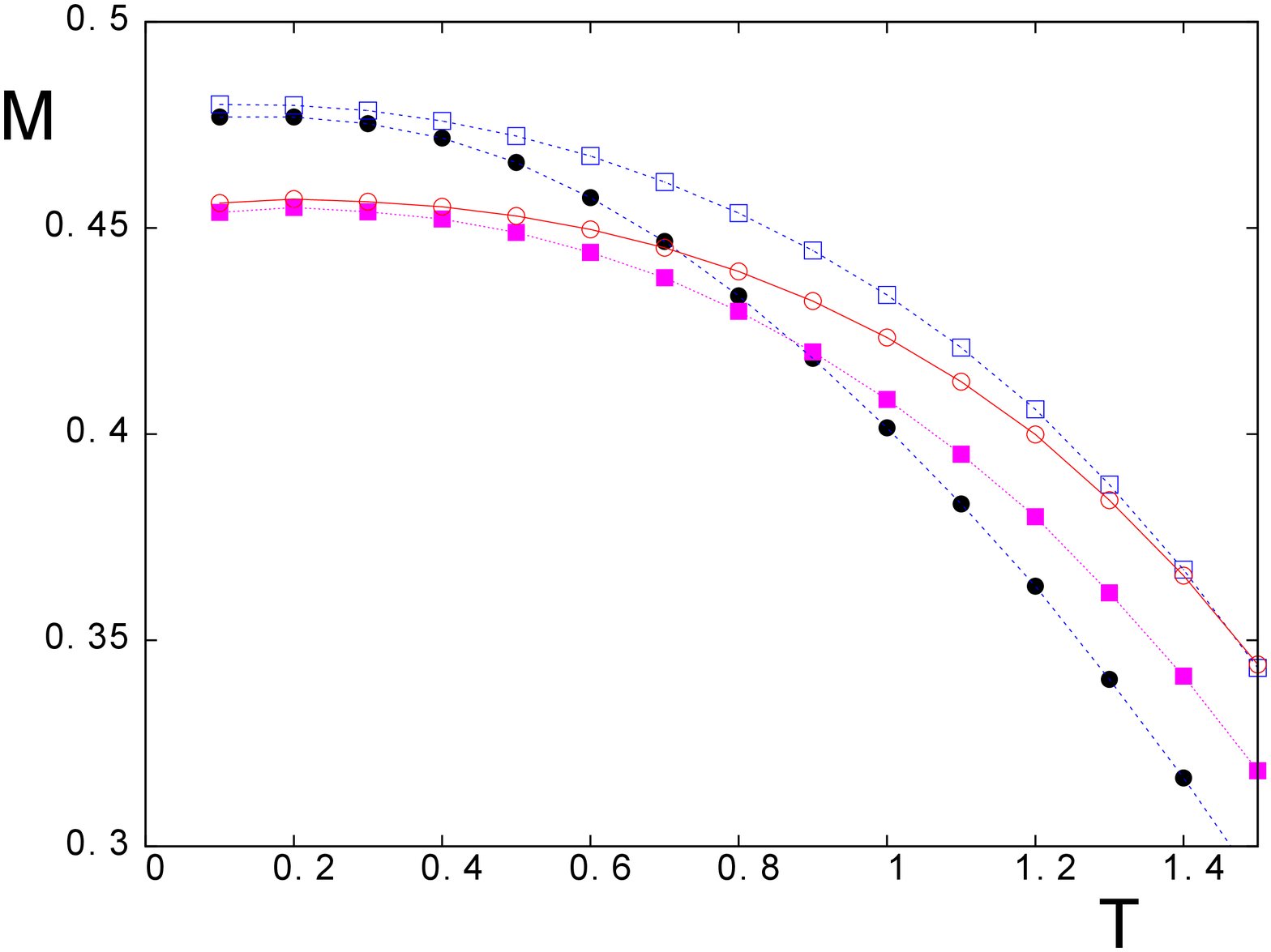}  % .eps
\caption{ (Color online) Layer magnetizations as functions of $T$ for $J_2/J_1=-2$ with $d=0.1$, $N_z=8$ (top). Zoom of the region at low $T$ to show crossover (bottom). Black circles, blue void squares, magenta squares and red void circles are for first, second, third and fourth layers, respectively. See text.}\label{magnet20}
\end{figure}

Let us show two examples of the magnetization, layer by layer, from the film surface in Figs. \ref{magnet14} and \ref{magnet20}, for the case where  $J_2/J_1=-1.4$ and -2 in a $N_z=8$ film.   Let us comment on the case $J_2/J_1=-1.4$:

(i) the shown result is obtained with a convergence of $1\%$. For temperatures closer to the transition temperature $T_c$, we have to lower the precision to a few percents which reduces the clarity because of their close values (not shown).

(ii) the surface magnetization, which has a large value at $T=0$ as seen in Fig. \ref{spin0}, crosses the interior layer magnetizations at $T\simeq 0.42$ to become much smaller than interior magnetizations at higher temperatures.  This crossover phenomenon is due to the competition between quantum fluctuations, which dominate low-$T$ behavior, and the low-lying surface spin-wave modes which strongly diminish the surface magnetization at higher $T$.  Note that the second-layer magnetization makes also a crossover at $T\simeq 1.3$.  Similar crossovers have been observed in quantum antiferromagnetic films \cite{DiepTF91} and quantum superlattices \cite{DiepSL89}.

Similar remarks can be also made for the case $J_2/J_1=-2$.

Note that though the layer magnetizations are different at low temperatures, they will tend to zero at a unique transition temperature as seen below.  The reason is that as long as an interior layer magnetization is not zero, it will act on the surface spins as an external field, preventing them to become zero.

\begin{figure}[htb]
\centering
\includegraphics[width=7cm,angle=0]{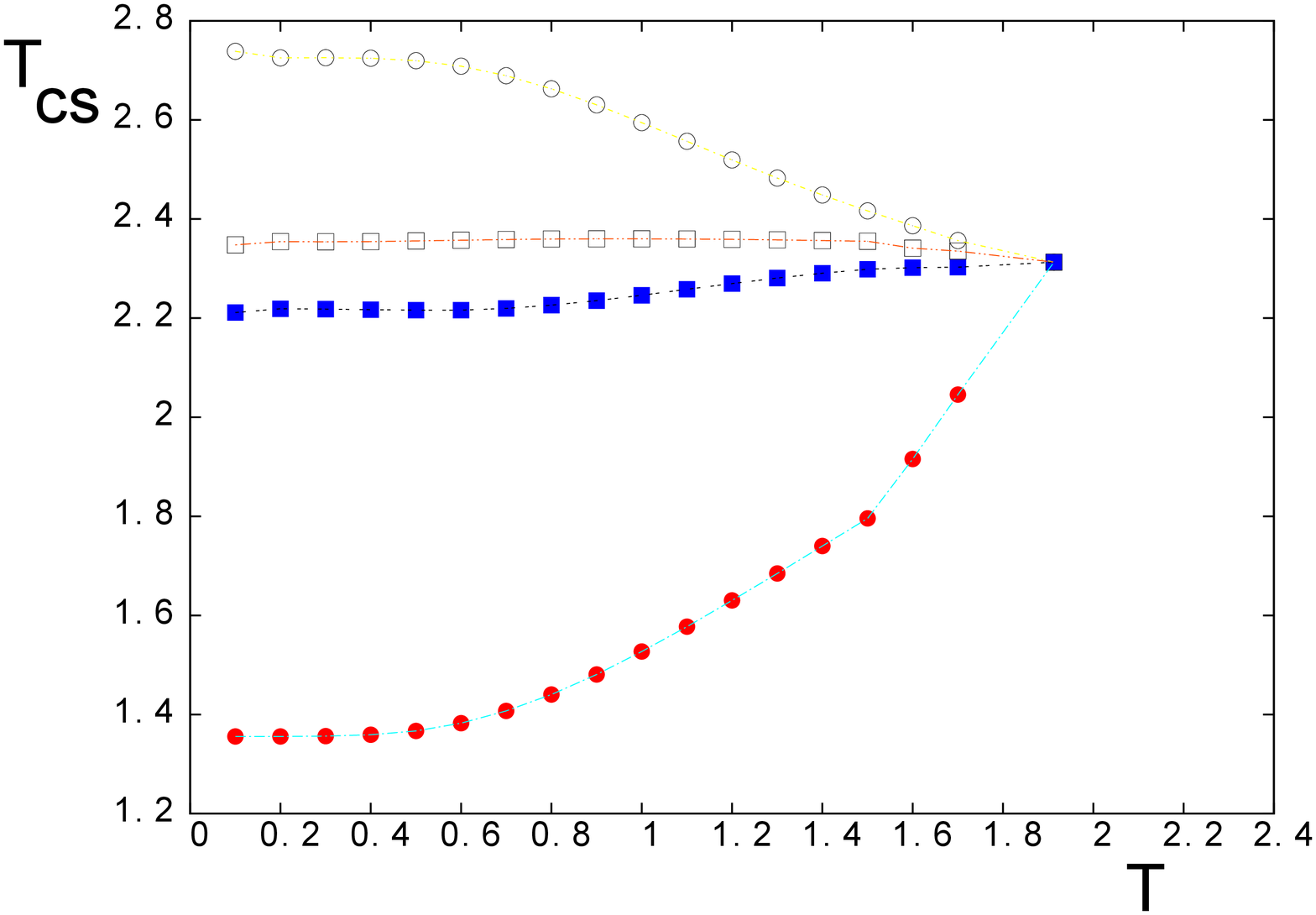}  % .eps
\includegraphics[width=7cm,angle=0]{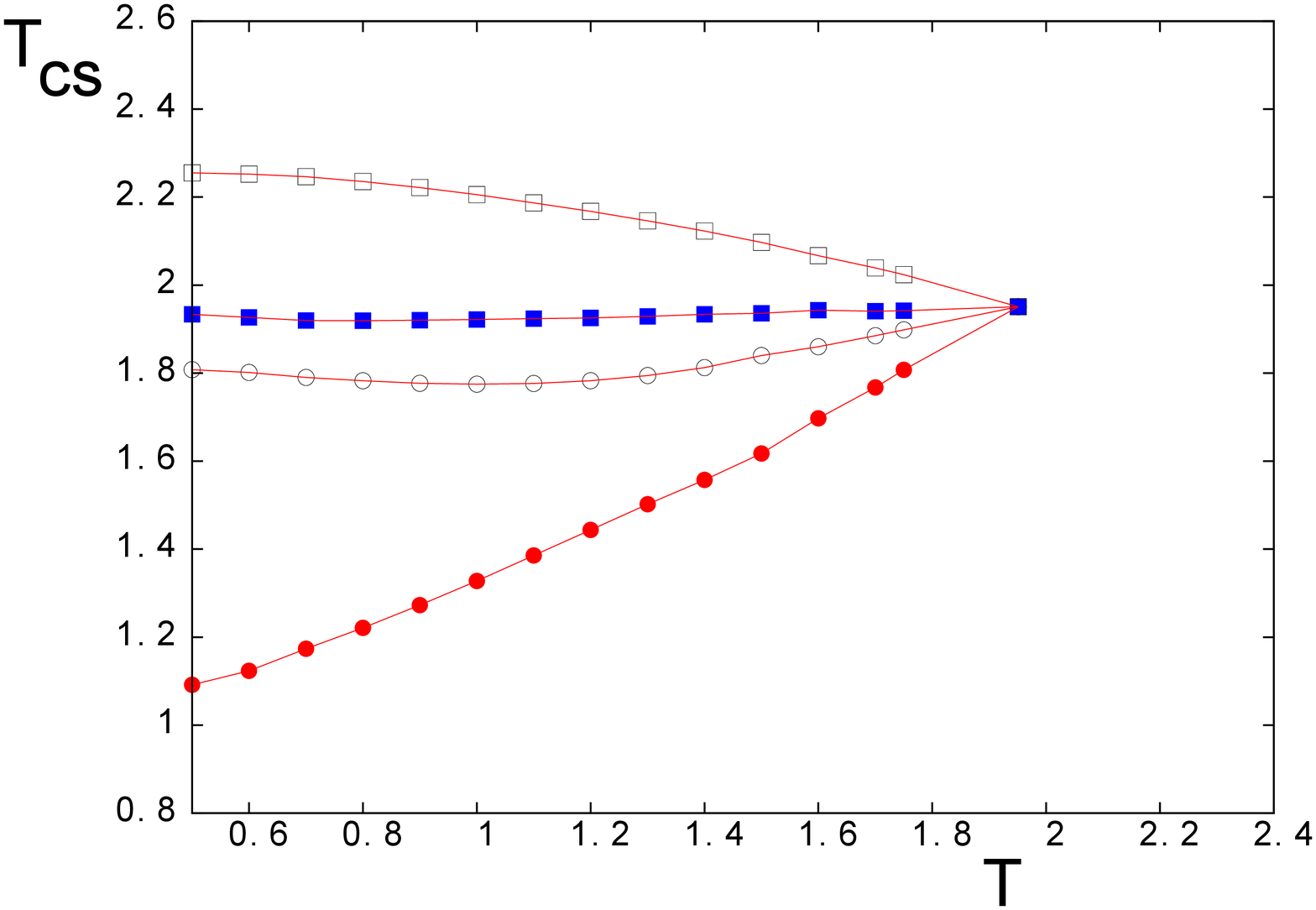}  % .eps
\caption{ (Color online) Top: Transition temperature is calculated at $p=J_2/J_1=-1.4$ for $d=0.1$, $N_z=8$:  at each temperature, using the self-consistent values of layer magnetizations at $T<T_c$, Eq.(\ref{tcc}) is solved to obtain  $T_{cs}$ for each layer.  The convergence is reached when $T_{cs}$ tend to a single value $T_c$. One has $T_c\simeq 2.313\pm 0.010$. Red circles, black void circles, blue squares and blue void squares are  $T_{cs}$ obtained from Eq. (\ref{tcc}) for first, second, third and fourth layers, respectively, at different temperatures. Bottom: Extrapolation by lines to obtain $T_c$ is shown for surface parameter $J_1^s/J_1=0.7$.  The precision for self-consistent convergence is $1\%$ for layer magnetizations. See text for comments. }\label{tc-conver}
\end{figure}

\begin{figure}[htb]
\centering
\includegraphics[width=7cm,angle=0]{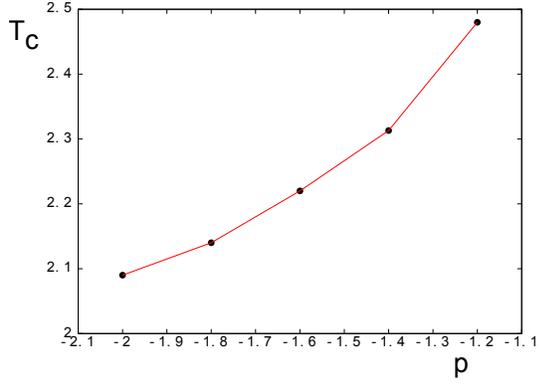}  % .eps
\caption{ (Color online) Transition temperature versus $p=J_2/J_1$ for an 8-layer film with $d=0.1$. See text for comments. }\label{figtc}
\end{figure}

The temperature where layer magnetizations tend to zero is calculated by Eq. (\ref{tcc}). Since all layer magnetizations tend to zero from different values, we have to solve self-consistently $N_z$ equations (\ref{tcc}) to obtain the transition temperature $T_c$. One way to do it is to use the self-consistent layer magnetizations obtained as described above at a temperature as close as possible to $T_c$ as input for Eq. (\ref{tcc}). As long as the $T$ is far from $T_c$ the convergence is not reached: we have four 'pseudo-transition temperatures' $T_{cs}$ as seen in Fig. \ref{tc-conver}, one for each layer.  The convergence of these $T_{cs}$ can be obtained by a short extrapolation from temperatures when they are rather close to each other. $T_c$ is thus obtained with a very small extrapolation error as seen  in Fig. \ref{tc-conver}  for $p=J_2/J_1=-1.4$: $T_c\simeq 2.313\pm 0.010$.
The results for several $p=J_2/J_1$ are shown in Fig. \ref{figtc}.

\subsection{Effect of anisotropy and surface parameters}
\begin{figure}[htb]
\centering
\includegraphics[width=7cm,angle=0]{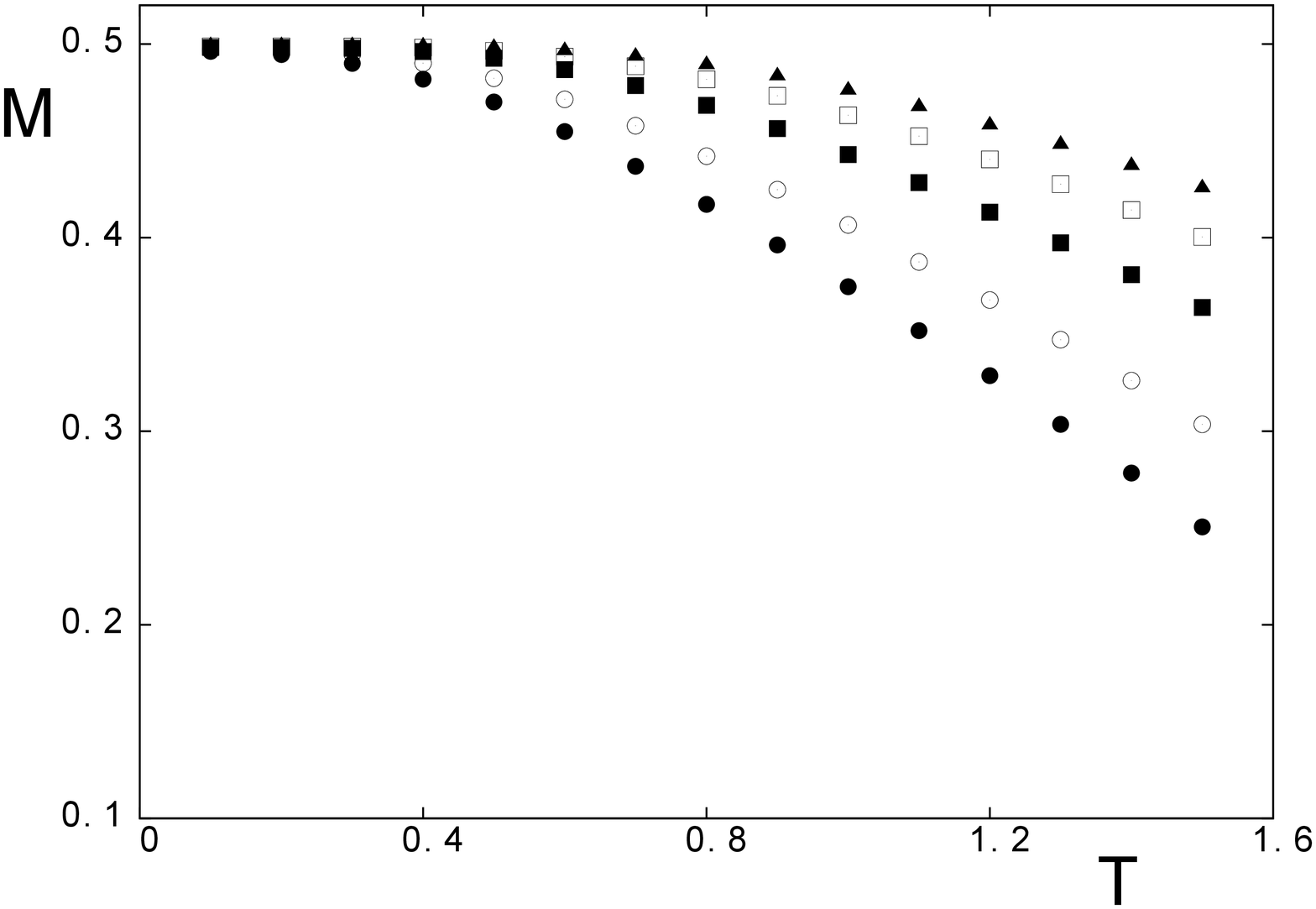}  % .eps
\caption{ Surface magnetization versus $T$ for $d=0.05$ (circles), 0.1 (void circles), 0.2 (squares), 0.3 (void squares) and 0.4 (triangles), with  $J_2/J_1=-1.4$,  $N_z=16$.}\label{figaniso}
\end{figure}

The results shown above have been calculated with an in-plane anisotropy interaction $d=0.1$.  Let us show now the effect of $d$. Stronger $d$ will enhance all the layer magnetizations and increase $T_c$. Figure \ref{figaniso} shows the surface magnetization versus $T$ for several values of $d$ (other layer magnetizations are not shown to preserve the figure clarity).
The transition temperatures are $2.091\pm 0.010$, $2.313\pm 0.010$, $2.752\pm 0.010$, $3.195\pm 0.010$ and $3.630\pm 0.010$ for $d=0.05$, 0.1, 0.2, 0.3 and 0.4, respectively.  These values versus $d$ lie on a remarkable straight line.

\begin{figure}[htb]
\centering
\includegraphics[width=7cm,angle=0]{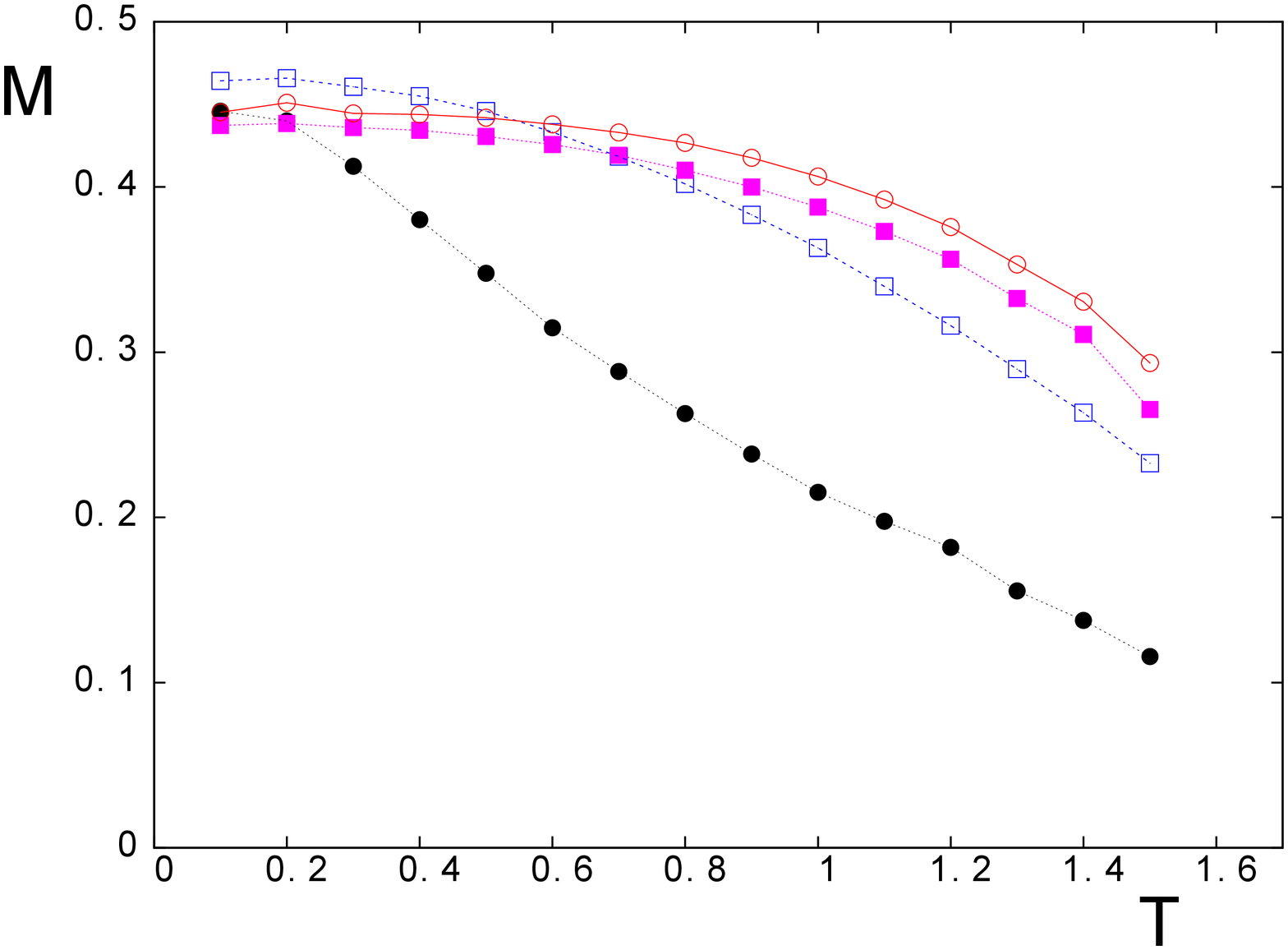}
\includegraphics[width=7cm,angle=0]{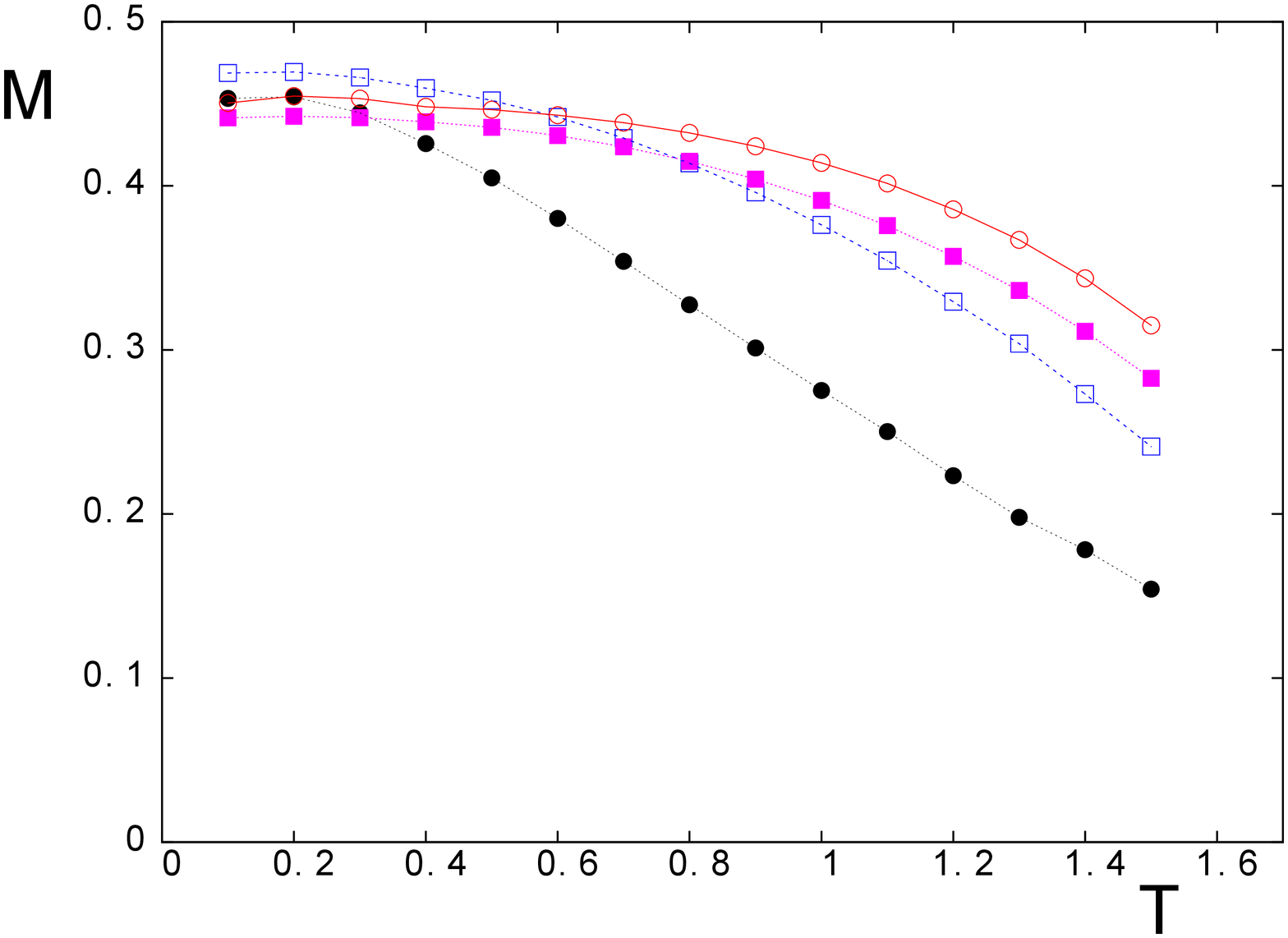}  % .eps
\includegraphics[width=7cm,angle=0]{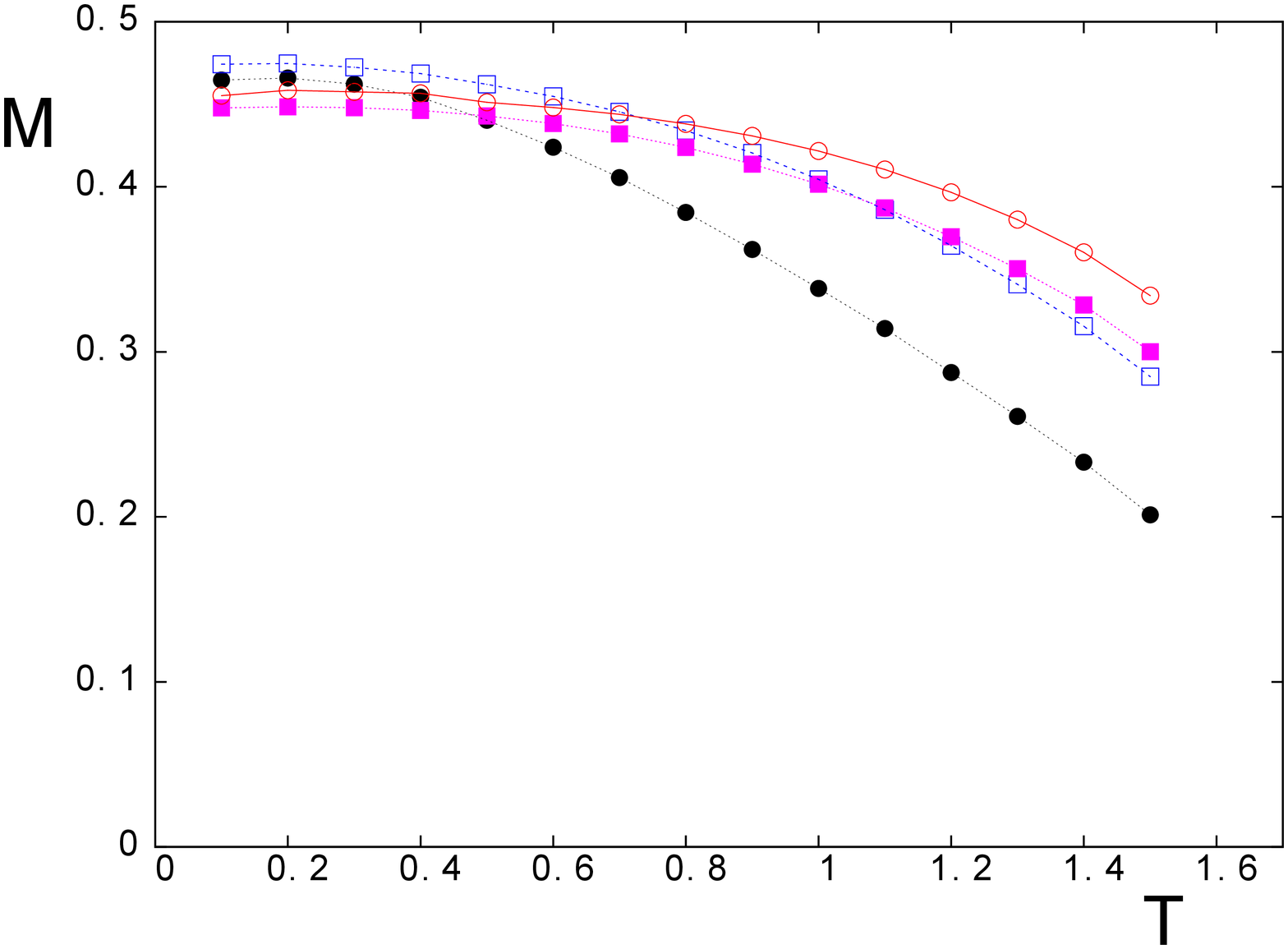}  % .eps
\caption{ (Color online) Layer magnetizations as functions of $T$ for the surface interaction $J_1^s=0.3$ (top), 0.5 (middle) and 0.7 (bottom) with $J_2/J_1=-2$, $d=0.1$ and $N_z=16$. Black circles, blue void squares, magenta squares and red void circles are for first, second, third and fourth layers, respectively.}\label{surf030507}
\end{figure}

\begin{figure}[htb]
\centering
\includegraphics[width=7cm,angle=0]{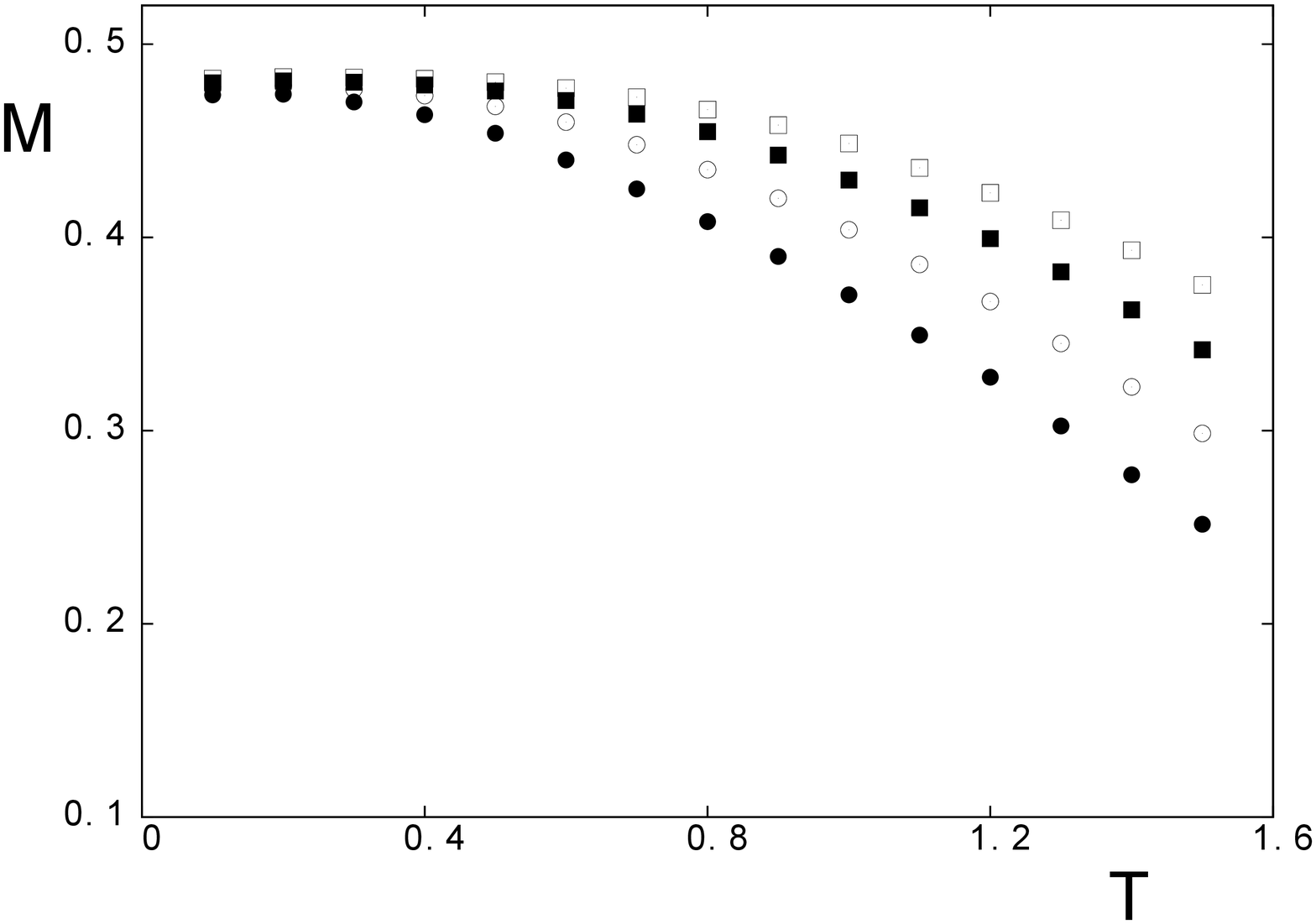}  % .eps
\includegraphics[width=7cm,angle=0]{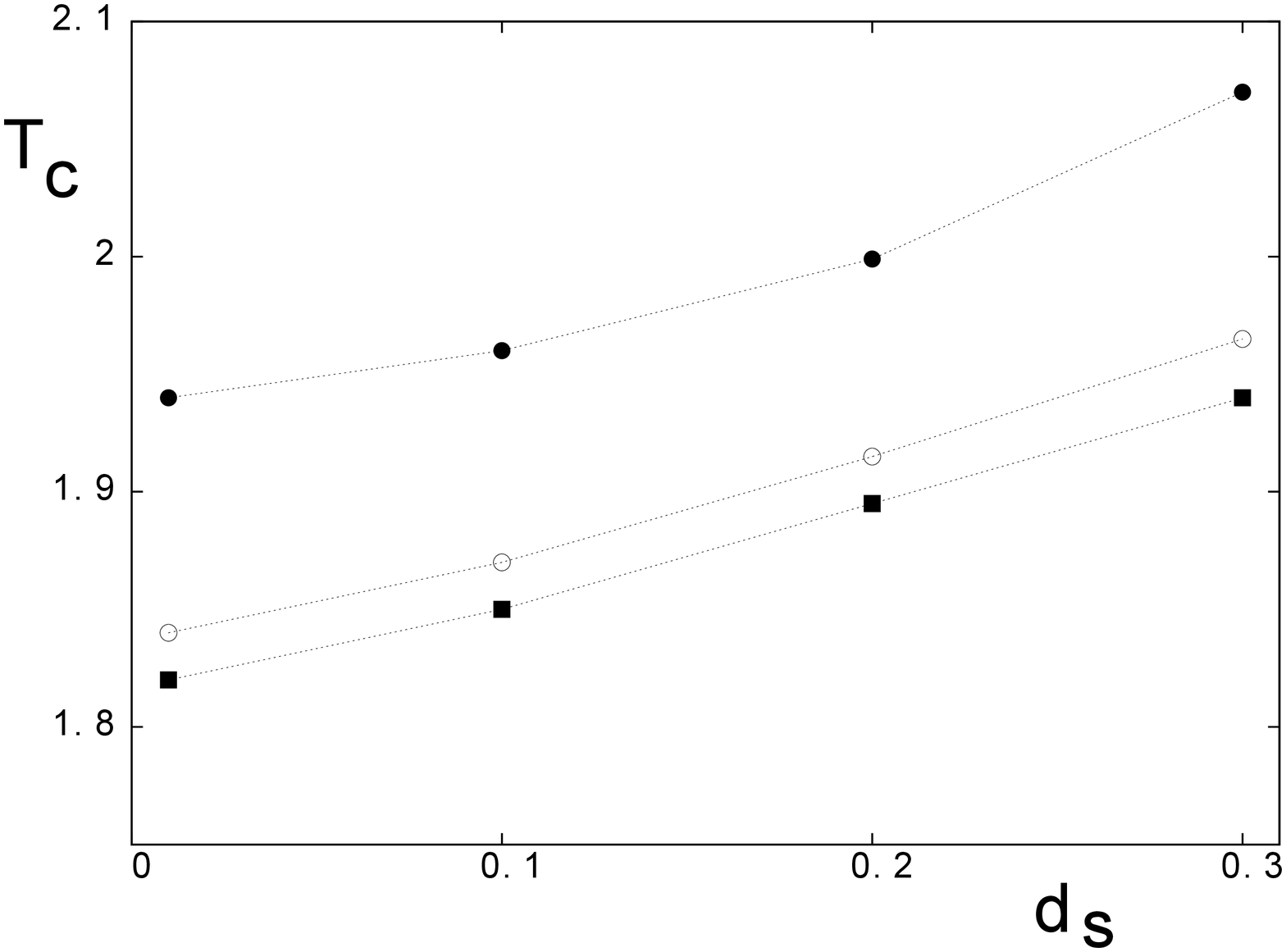}  % .eps
\caption{ Top: Surface magnetization versus $T$ for $d_s=0.01$ (circles), 0.1 (void circles), 0.2 (squares) and 0.3 (void squares), with $J_1^s=1$,  $J_2/J_1=-2$ and  $N_z=16$.  Bottom: Transition temperature versus $d_s$ for $J_1^s$=0.7, 0.5 and 0.3 (curves from up to down), with $J_2/J_1=-2$, $d=0.1$ and  $N_z=16$. }\label{figds}
\end{figure}

Let us examine the effects of the surface anisotropy and exchange parameters $d_s$ and $J_1^s$.  As seen above, even in the case where the surface interaction parameters are the same as those in the bulk the surface spin-wave modes exist in the spectrum. These localized modes cause a low surface magnetization observed in Figs. \ref{magnet14} and \ref{magnet20}.  Here, we show that with a weaker NN exchange interaction between surface spins and the second-layer ones, namely $J_1^s<J_1$, the surface magnetization becomes even much smaller with respect to the magnetizations of interior layers. This is shown in Fig. \ref{surf030507} for several values of $J_1^s$.  We observe again here the crossover of layer magnetizations at low $T$ due to quantum fluctuations as discussed earlier.  The transition temperature strongly decreases with $J_1^s$: we have $T_c=2.103\pm 0.010$,  $1.951\pm 0.010$, $1.880\pm 0.010$ and $1.841\pm0.010$ for $J_1^s=1$, 0.7, 0.5 and 0.3, respectively ($N_z=16$, $J_2/J_1=-2$, $d=d_s=0.1$).
Note that the value $J_1^s=0.5$ is a very particular value: the GS configuration is a uniform configuration with all angles equal to  $60^\circ$, namely there is no surface spin rearrangement.  This can be explained if we look at the local field acting on the surface spins: the lack of neighbors is compensated by this weak positive value of $J_1^s$ so that their local field is equal to that of a bulk spin. There is thus no surface reconstruction. Nevertheless, as $T$ increases, thermal effects will strongly diminish the surface magnetization as seen in Fig. \ref{surf030507} (middle).  As for the surface anisotropy parameter $d_s$, it affects strongly the layer magnetizations and the transition temperature: we show in Fig. \ref{figds} the surface magnetizations and the transition temperature for several values of $d_s$.

\subsection{Effect of the film thickness}
We have performed calculations for $N_z=8$, 12 and 16. The results show that the effect of the thickness at these values is not significant: the difference lies within convergence errors.
 Note that the classical ground state of the first four layers is almost the same: for example, here are the values of cosinus of the angles of the film first half for $N_z=16$ which are to be compared with the values for $N_z=8$ given in Table I, for $p=J_2/J_1=-2$ (in parentheses are angles in degree):

\noindent 0.86737967 (29.844446),  0.41125694 (65.716179), 0.52374715 (58.416061), 0.49363765 (60.420044),
 0.50170541 (59.887100),  0.49954081 (60.030373),  0.50013113 (59.991325),  0.49993441 (60.004330).

\noindent From the 4th layer, the angle is almost equal to the bulk value ($60^\circ$).

At $p=J_2/J_1=-2$, the transition temperature is $2.090\pm 0.010$ for $N_z=8$, $2.093\pm 0.010$ for $N_z=12$ and $2.103\pm 0.010$ for $N_z=16$. These are the same within errors.  At smaller thicknesses, the difference can be seen. However, for helimagnets in the $z$ direction,  thicknesses smaller than 8 do not allow to see fully the surface helical reconstruction which covers the first four layers: to study surface helical effects in such a situation would not make sense.

At this stage, it is interesting to note that our result is in excellent agreement with experiments: it has been experimentally observed that the transition temperature  does not vary significantly in MnSi films in the  thickness range of $11-40$ nm \cite{Karhu2011}.  One possible explanation is that the helical structure is very stable as seen above: the surface perturbs the bulk helical configuration only at the first four layers, so the bulk 'rigidity' dominates the transition. This has been experimentally seen in holmium films \cite{Leiner}.

 \begin{figure}[htb]
\centering
\includegraphics[width=7cm,angle=0]{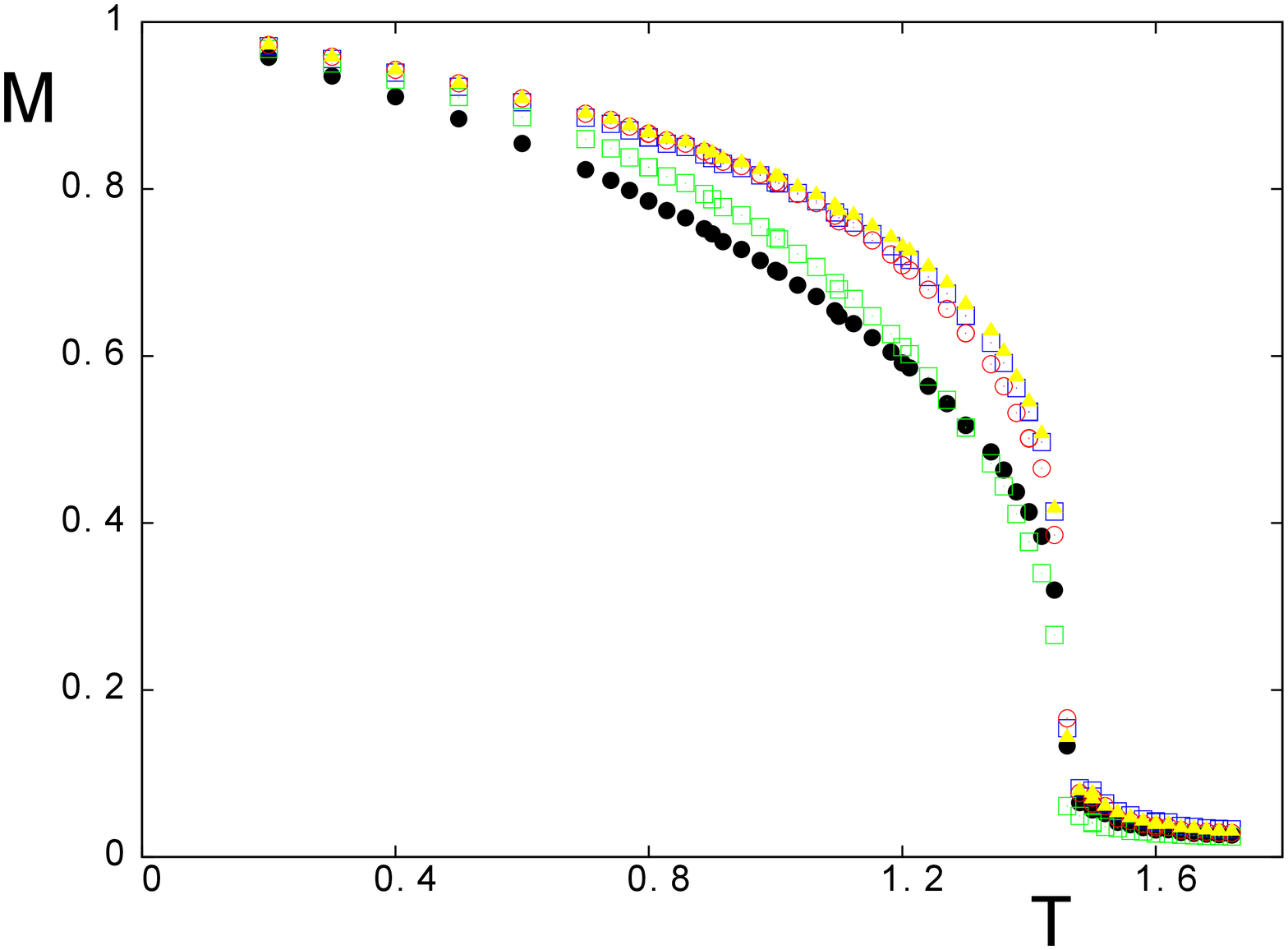}
\includegraphics[width=7cm,angle=0]{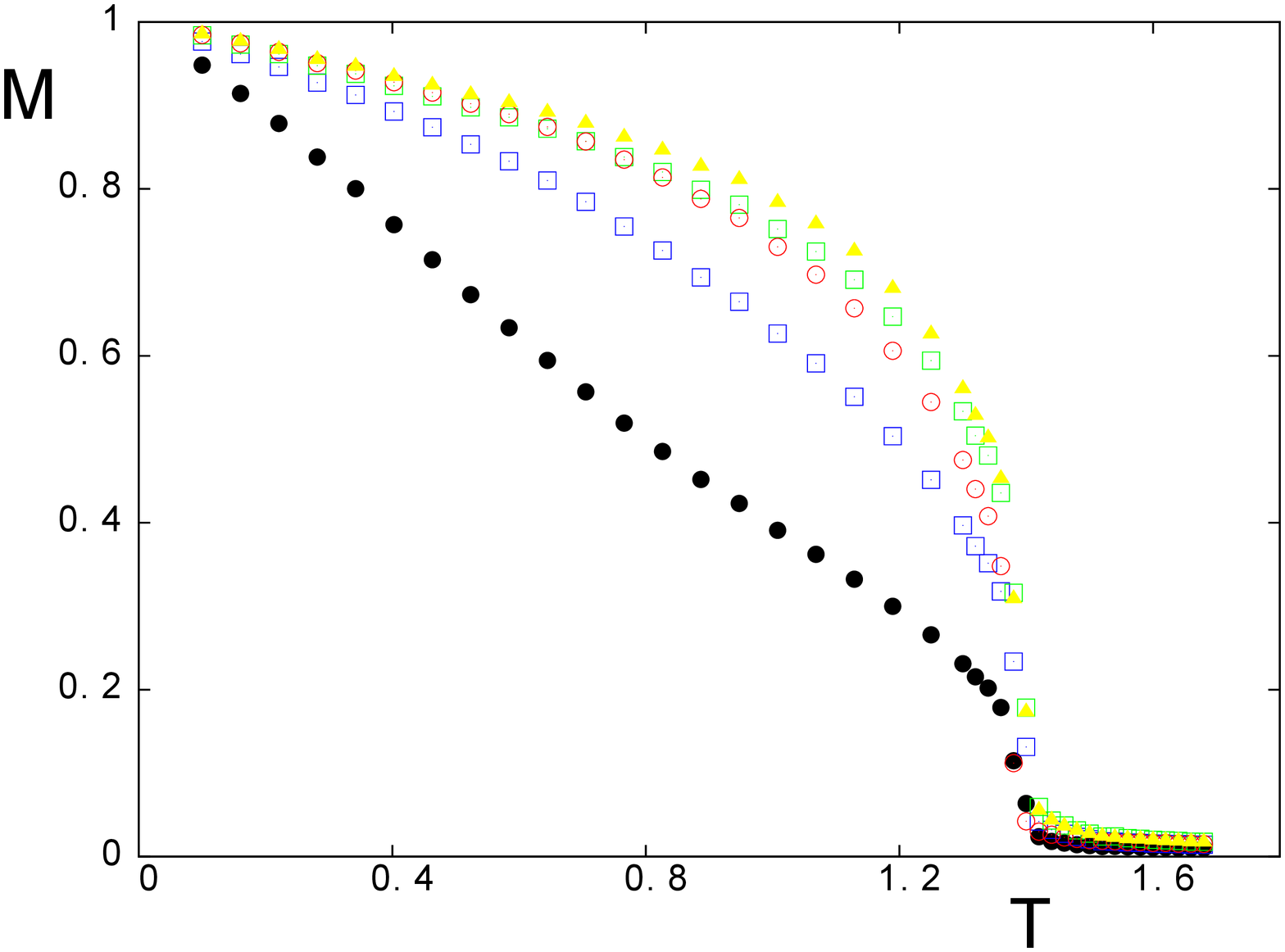}  % .eps
\caption{ (Color online) Monte Carlo results: Layer magnetizations as functions of $T$ for the surface interaction $J_1^s=1$ (top) and 0.3 (bottom) with $J_2/J_1=-2$ and $N_z=16$. Black circles, blue void squares, cyan squares and red void circles are for first, second, third and fourth layers, respectively.}\label{surf03MC}
\end{figure}

\subsection{Classical helimagnetic films: Monte Carlo simulation}

To appreciate quantum effects causing crossovers of layer magnetizations presented above at low temperatures, we show here some results of the classical counterpart model: spins are classical XY spins of amplitude $S=1$. We take the XY spins rather than the Heisenberg spins for comparison with the quantum case because in the latter case we have used an in-plane Ising-like anisotropy interaction $d$.  Monte Carlo simulations have been carried out over film samples of $100\times 100\times 16$. Periodic boundary conditions are applied in the $xy$ plane.  One million of MC steps are discarded to equilibrate the system and another million of MC steps are used for averaging. The layer magnetizations versus $T$ are shown in Fig. \ref{surf03MC} for the case where surface interaction $J_1^s=1$ (top) and 0.3 (bottom) with $J_2/J_1=-2$ and $N_z=16$.
One sees that i) by extrapolation there is no spin contraction at $T=0$  and there is no crossover of layer magnetizations at low temperatures, ii) from the intermediate temperature region up to the transition the relative values of layer magnetizations are not always the same as in the quantum case: for example at $T=1.2$, one has $M_1<M_3<M_4<M_2$ in Fig. \ref{surf03MC} (top) and  $M_1<M_2<M_4<M_3$ in Fig. \ref{surf03MC} (bottom) which are not the same as in the quantum case shown in Fig. \ref{magnet20} (top) and Fig. \ref{surf030507} (top).  Our conclusion is that even at temperatures close to the transition, helimagnets may have slightly different behaviors according to their quantum or classical nature.  Extensive MC simulations with size effects and detection of the order of the phase transition is not the scope of this present paper.

 \section{Conclusion}\label{conclu}

 We have studied in this paper surface effects in a helimagnet of body-centered cubic lattice with quantum Heisenberg spins.
 The classical bulk ground-state spin configuration is exactly calculated and is found to be strongly modified near the film surface. The surface spin rearrangement is however limited to the first four layers in our model, regardless of the bulk angle, namely the NNN interaction strength $J_2$. The spin-wave excitation is calculated
 using a general Green's function technique for non collinear spin configurations. The layer magnetization as a function of temperature as well as the transition temperature are shown for various interaction parameters.  Among the striking features found in the present paper, let us mention i) the cross-over of layer magnetizations at low temperatures due to the competition between quantum fluctuations and thermal effects, ii) the existence of low-lying surface spin-wave modes which cause a low surface magnetization, iii) a strong effect of the surface exchange interaction ($J_1^s$) which drastically modifies the surface spin configuration and gives rise to a very low surface magnetization, iv) the transition temperature varies strongly with the helical angle but it is insensitive to the film thickness in agreement with experiments performed on MnSi films \cite{Karhu2011} and holmium \cite{Leiner}, v) the classical spin model counterpart gives  features slightly different from those of the quantum model, both at low and high temperatures.

 To conclude, let us emphasize that the general theoretical method proposed here allows us to study at a microscopic level surface spin-waves and their physical consequences at finite temperatures in systems with non collinear spin configurations such as helimagnetic films.  It can be used in more complicated situations such as helimagnets with Dzyaloshinskii-Moriya interactions \cite{Karhu2012}.

{}

\end{document}